\documentclass[preprint,3p]{elsarticle}

\usepackage[pdftex]{hyperref}
\hypersetup{
    colorlinks   = true, 
    urlcolor     = blue, 
    linkcolor    = blue, 
    citecolor   = red 
}
\usepackage[labelformat=simple]{subcaption}
\usepackage{graphics}
\usepackage{epsfig}

\usepackage{float}
\usepackage{amssymb}
\usepackage{amsmath}
\usepackage{amsthm}
\usepackage{amsfonts}
\usepackage{dsfont}

\usepackage{appendix}

\usepackage{algorithm}
\usepackage{algorithmic}
\usepackage{xcolor}
\usepackage[noabbrev]{cleveref}

\usepackage{mathrsfs}

\newcommand{\Vn}{\mathbf{n}}

\newcommand{\Par}{\mathscr{P}}

\newcommand{\Sk}{\mathscr{S}}

\journal{IJNME}

\begin{document}
    
\begin{frontmatter}
    
\title{Adaptive stabilized finite elements: Continuation analysis of compaction banding in geomaterials}

\author[label1,label2]{Roberto J. Cier\corref{cor1}}
\cortext[cor1]{Corresponding Author}
\ead{rcier93@gmail.com}
\author[label2]{Thomas Poulet}
\author[label3]{Sergio Rojas}
\author[label2,label3]{Victor M. Calo}
\author[label4]{Manolis Veveakis}

\address[label1]{School of Civil and Mechanical Engineering, Curtin University, Kent Street, Bentley, Perth, WA 6102, Australia}
\address[label2]{CSIRO  Mineral Resources, 26 Dick Perry Avenue, Kensington, WA 6151, Australia}
\address[label3]{School of Earth and Planetary Sciences, Curtin University, Kent Street, Bentley, Perth, WA 6102, Australia}
\address[label4]{Civil and Environmental Engineering, Duke University, Durham, NC 27708-0287, USA}

\begin{abstract}
Under compressive creep, visco-plastic solids experiencing internal mass transfer processes have been recently proposed to accommodate singular cnoidal wave solutions, as material instabilities at the stationary wave limit. These instabilities appear when the loading rate is significantly faster than the capability of the material to diffuse internal perturbations and lead to localized failure features (e.g., cracks and compaction bands). This type of solution, generally found in fluids, has strong nonlinearities and periodic patterns. Due to the singular nature of the solutions, the applicability of the theory is currently limited. Additionally, effective numerical tools require proper regularization to overcome the challenges that singularity induces. We focus on the numerical treatment of the governing equation using a nonlinear approach building on a recent adaptive stabilized finite element method. This method provides a residual representation to drive adaptive mesh refinement, a particularly useful feature for the problem at hand. We compare against analytical and standard finite element solutions to demonstrate the performance of our approach. We then investigate the sensitivity of the diffusivity ratio, main parameter of the problem, and identify multiple possible solutions, with multiple stress peaks. Finally, we show the evolution of the spacing between peaks for all solutions as a function of that parameter.
\end{abstract}


\end{frontmatter}

\section{Introduction}\label{sec:introduction}
Localization phenomena play a predominant role in Earth Sciences as many interesting geological features fall within this category, including faults, folds, boudinage, landslides, and mineralization, just to name a few. Among all localized geological features, spatially periodic patterns are increasingly gaining popularity due to their impact in a wide range of applications, such as mining~\cite{Iophis2007}, particularly now when larger-scale and deeper mines are pursued. Localization features are particularly relevant when they affect permeability, which plays a critical role in various fields including mineral exploration~\cite{Hayward2018, Kelka2017} and unconventional resources exploration~\cite{Regenauer-Lieb2016} as well as their exploitation.

One mechanism responsible for spatially periodic patterns affecting porosity and permeability is the formation of compaction bands. These bands are narrow flat zones of deformation perpendicular to the maximum compressive principal stress~\cite{Das2013}. For example, a succession of compacted zones of lower permeability in a non-compacted higher permeability background is the most intuitive way to imagine compaction bands~\cite{Holcomb2003}, with critical implications on the fluid flow in terms of creation of impermeable barriers as well as compartmentalized reservoirs and aquifers. These compaction bands therefore have significant impact on the fluid production or geological storage (${\text{CO}_2}$, nuclear waste). Several models exist to describe the mechanism on its in situ periodic occurrence, e.g.~by Cecinato~\cite{Cecinato2014}, however, there also exist other physical processes that lead to periodic bands under compression with increased permeability, an example from the melt segregation field known as decompaction bands~\cite{Rabinowicz2004}. 

Attempts to theoretically describe deformation bands (principally of shear-type) start early in the theory of plasticity~\cite{Hill1950}. For geomaterials, this analysis started with the work of Rudnicki and Rice~\cite{Rudnicki1975}, who extensively study the conditions for localized deformation in low porosity rocks using bifurcation analysis. Although that theory sought to explain the appearance of shear bands in a specific type of rocks, a subsequent work by Olsson~\cite{Olsson1999} confirmed that the origin of compaction bands could also be explained within this framework. Later, a re-examination of the original theory was developed by Issen and Rudnicki~\cite{Issen2001}, where the triggering conditions of both shear and compaction bands could be deduced for a broader range of materials. Thus, they proposed the introduction of a cap-type yield function to take into account compactive yield in high porosity rocks. Finally, in the particular case of cemented soils, studies on the onset of compaction bands were developed by Arroyo et al.~\cite{Arroyo2005} using a bonded soil model and bifurcation conditions established by Rudnicki and Rice~\cite{Rudnicki1975}.

From a numerical perspective, little has been done on the simulation of compaction bands compared with the numerous studies on the formation of shear banding~\cite{Oka2011}. Within the most relevant works can be named the computational modeling carried out by Borja~\cite{borja2004}, where the onset conditions for deformation bands were analyzed using the single hardening constitutive model proposed by Kim and Lade~\cite{Kim1988}. In addition to this, Oka et al.~\cite{Oka2011} developed finite element simulations in diatomaceous mudstone using an elasto-viscoplastic model and compared them with triaxial tests results. Although their models predicted compaction bands for higher confining pressures, they did not tackle the challenging problems of identifying the onset conditions of the phenomenon and its periodicity for a broader range of confinement stresses. Other experimental studies were also developed for sandstone specimens~\cite{Fortin2006} and calcarenite samples~\cite{Baxevanis2006}.

All previous studies only focused on the mechanical problem, without considering multi-physical scenarios that could allow capturing other contributions to the deformation process. To overcome this issue, recently, Veveakis and Regenauer-Lieb~\cite{Veveakis2015} developed a wave mechanics approach which showed that regularly spaced localization bands of hydro-mechanical nature can appear in rocks under compaction. The wave description is radically different as this type of localized deformation stems from a generalization of Terzaghi's linear theory of consolidation to materials with non-linear visco-plastic rheology and any arbitrary type of internal mass transfer mechanism. Through these genearlizations, the authors derived the following governing equation for nonlinear consolidation, admitting material instabilities presented in the effective stress $\sigma’$ (for elasto-viscoplastic materials subject to hydro-mechanical processes):

\begin{equation}\label{eq:cnoidal_orig}
\frac{\partial^2\sigma'}{\partial{z^2}} - \lambda\sigma'^m =0
\end{equation}
In this dimensionless equation $\lambda$ represents a ratio between the particular diffusive processes of the problem (i.e., the mechanical deformation of the matrix and the internal mass exchange) and $m$ is a material-dependent pressure exponent. Under certain conditions, the solution of~\eqref{eq:cnoidal_orig} presents numerical instabilities due to the loading rate being faster than the mass diffusion rate; thus mass variations in the specimen cannot be equilibrated, producing zones of stress concentrations that represent the compaction bands. This phenomenon produces periodic volumetric failure patterns denoted as cnoidal waves~\cite{Regenauer-Lieb2016,Veveakis2015,Veveakis2014}, due to the analogy with its counterpart in fluids dynamics. “Cnoidal wave” is the term used for the solution of the Korteweg-de Vries (KdV) equation~\cite{Korteweg1895} expressed regarding the square Jacobi $cn$ elliptic function. The KdV equation describes a traveling wave in shallow water surfaces, first observed by Russell~\cite{Russell1844}, which has been extensively used for many physical problems related to wave mechanics. For the case of solids, this cnoidal wave approach allows us to conclude that periodic instabilities are produced as a volumetric response during failure and that the latter is controlled by deformation rate, and not by critical stress or hardening reaching as stated by classical theories~\cite{Veveakis2015}.

While the original theory~\cite{Veveakis2015} captures well the essence of hydro-mechanical instabilities, its simplifying nature leads to unbounded stress values, which are not realistically possible. A complementary study~\cite{Alevizos2017} considered, for instance, the effects of chemical reactions, which introduce a regularization term in the equation and allow the effective stress to remain capped. Without entering the debate about which physical processes could occur on the back of stress peaks, we find if convenient to consider a nonlinear regularization term $N_r(\sigma')$ for mathematical reasons. This regularization allows us to deal with bounded continuous solutions. As such, we consider a generalization of equation~\eqref{eq:cnoidal_orig} of the form
\begin{equation}\label{eq:cnoidal_chem}
\frac{\partial^2\sigma'}{\partial{z^2}} - \lambda\sigma'^m + N_r(\sigma') = 0,
\end{equation}
and for this study, we use the existing formulation of~\cite{Alevizos2017}. We succinctly recall it in Appendix~\ref{AppxA} for completeness.

In this work, we extend an adaptive stabilized finite element method (FEM) to overcome the numerical problems by which stress singularities hinder the effectiveness and robustness of numerical schemes. We showcase the power of such approach by solving this equation consistently. Next, we implement an arc-length continuation algorithm to perform a numerical bifurcation analysis to better understand the parameter sensitivity of the model. This analysis allows us to derive important conclusions regarding the stability regimes of the system. We introduce the numerical approach used for the resolution of~\eqref{eq:cnoidal_chem} in Section~\ref{sec:num} and all the numerical outcomes in Section~\ref{sec:numexp_1d}, leading to the discussion in Section~\ref{sec:discussion}.

\section{Numerical approximation of the cnoidal problem}\label{sec:num}
The cnoidal wave approach in solids seeks to explain the formation of specific localized deformation bands more readily than alternative explanations provided by classical theories. While the cnoidal approach offers a new perspective to the localization phenomenon, there are some points related to the solution of its governing equation that need to be addressed before a detailed study is possible. Equation~\eqref{eq:cnoidal_chem} has known analytical solutions only for the integer values of $m=1, 2, 3$. However, solutions for higher or non-integer values of $m$ need numerical treatment, and, to date, there has been no successful attempt to numerically solve this equation satisfactorily. The lack of a robust numerical solution for this equation is related to the complexity of the treatment of this class of nonlinear problems. We seek to overcome this issue by developing a consistent numerical solution for~\eqref{eq:cnoidal_chem}. Therefore, we use the new adaptive stabilized finite element method based on residual minimization, developed by Calo et al.~\cite{calo2019}, and extend its application to nonlinear problems. We opt for using this formulation instead of rather than alternative FEM approximations because of the stability properties this method enjoys, and its built-in adaptive mesh refinements, a crucial feature for localization of instabilities. 

\subsection{Weak variational formulation}
In an abstract setting, we consider a well-posed variational formulation for a general nonlinear problem. For an open set $\Omega\neq0$ with boundary $\partial \Omega$, and Hilbert spaces $U$ (trial) and $V$ (test), let $N$ be a differentiable nonlinear map with Fr\'echet derivative $DN(u)$ at ${u \in \Omega}$. We associate the nonlinear map ${n:U\times V \rightarrow \mathbb{R}, \, n(u;v):=\langle N(u),v \rangle}$, where ${\langle \boldsymbol{\cdot}, \boldsymbol{\cdot} \rangle}$ represents the duality pairing in $V$. Let ${n'(u;z,v)}$ abbreviate the derivative $DN(u)$, around a known value $u$, and in the direction of an increment $z$:
\begin{equation}\label{eq:der-def}
 n'(u;z,v):=\langle DN(u;z),v\rangle = \frac{d}{d\epsilon}n(u+\epsilon z; v)\big|_{\epsilon=0} \quad \text{for } u \in \Omega, z \in U, v \in V.
\end{equation}
We finally set $\ell(\boldsymbol{\cdot}): V \rightarrow \mathbb{R}$ as a continuous linear form. Hence, the weak formulation for a nonlinear problem reads:
\begin{equation}\label{eq:abs_vf}
\left\{
\begin{array}{l}
\text{Find } u \in U, \text{ such that:} \smallskip \\
\begin{array}{l}
n(u;v)= \ell(v), \quad \forall v \in V,
\end{array}
\end{array}
\right.
\end{equation}

We now state the variational formulation for the particular case of the cnoidal equation~\eqref{eq:cnoidal_chem}. Following~\cite{Veveakis2015}, we define the boundary conditions as $\sigma' = 1 $ on $\partial \Omega$. In order to derive the continuous formulation of~\eqref{eq:cnoidal_chem}, we split ${\sigma'=u+1}$, and the steady state equation with homogeneous Dirichlet boundary conditions then reads:
\begin{equation}\label{eq:numsol_def_equation}
\left\{\begin{array}{l}
\text{Find } u=\sigma' -1 \text{ such that:} \smallskip \\
\begin{array}{rcl}
\Delta u-\mathcal{F}(u)&=&0 \quad \text{in} \quad \Omega,\\
u&=&0 \quad \text{on} \quad \partial \Omega,
\end{array}
\end{array}\right.
\end{equation} 
with ${\mathcal{F}(u)=\lambda\, (1+u)^m -\mu \exp(\beta u)}$. In~\eqref{eq:numsol_def_equation}, ${\mu \exp(\beta u)}$ is equivalent to the regularization term $N(\sigma')$ from~\eqref{eq:cnoidal_chem}, which avoids the unbounded stress growth of the exponent ${m>1}$ capping the value of $u$ to finite values and making $\mathcal{F}(u)$ to remain positive. 

\sloppy We consider the standard notation for Hilbert spaces: ${L^2(\Omega)=\{w: \Omega\in \mathbb{R}^d  \rightarrow \mathbb{R} : \int_{\Omega}|w|^2<+\infty\}}$,  ${H^1(\Omega)=\{w\in L^2(\Omega): \nabla w \in [L^2(\Omega)]^d\}}$ and ${H^1_0(\Omega)=\{w\in H^1(\Omega): w=0 \; \text{on} \; \partial\Omega\}}$. Multiplying~\eqref{eq:numsol_def_equation} by a test function $v$ and integrating by parts, we obtain the following weak variational formulation:
\begin{equation}\label{eq:weak_nonlin}
\left\{\begin{array}{l}
\text{Find } u \in H^1_0(\Omega), \text{ such that:} \smallskip \\
\begin{array}{l}
n(u;v)= \ell(v), \quad \forall v \in H^1_0(\Omega),
\end{array}
\end{array}\right.
\end{equation}
with $n(u;v)=(\nabla u, \nabla v)_{0, \Omega} + (\mathcal{F}(u), v)_{0, \Omega}$ and $\ell(v)=(f,v)_{0, \Omega}$, where $(\boldsymbol{\cdot}, \boldsymbol{\cdot})_{0, \Omega}$ represents the $L^2$ scalar product in $\Omega$. 

\subsection{Discontinuous Galerkin discretization}
In this section we briefly discuss a Discontinuous Galerkin (dG) formulation associated with~\eqref{eq:weak_nonlin} which allows us to construct the adaptive stabilized formulation. We use as a starting point the dG discretization based on classical interior penalty schemes for elliptic problems~\cite{ern2006, cockburn2012}.
\subsubsection{Discrete setting}
Let $\{\Par_h\}$ be a family of simplicial meshes of $\Omega$. For simplicity, we assume that $\Omega$ is exactly represented by any mesh in $\Par_h$, that is, $\Omega$ is an interval, a polygon, or a polyhedron. We denote $T$ as the generic element in $\Par_h$, with boundary $\partial T$, diameter $h_T$, and unit outward normal $\Vn_T$. We set $h = \max_{T \in \Par_h} h_T$ and we assume, without loss of generality, that $h \leq 1$. We define the classical dG approximation space
\begin{align}
V_h :=\{v_h \in L^2(\Omega) \ | \  \forall T \in \Par_h, v_h|_T \in \mathds{P}_k \},
\end{align}
where $\mathds{P}_k$ denotes the set of polynomials, defined over $T$, with polynomial degree smaller or equal than $k$. It is also convenient to set the extended space ${V_{h,\#} =H^2(\Par_h)+V_h}$.

We collect all the faces or edges $F$ of $\Omega_h$ into the set ${\Sk_h= \bigcup_{T\in\Par_h} F}$. We define the boundary skeleton $\Sk^\partial_h$ as ${\Sk^\partial_h = \Sk_h \cap \Gamma}$, and the internal skeleton $\Sk^0_h$ as ${\Sk^0_h=\Sk_h \backslash \Gamma}$.  Over $\Sk_h$, we define $\Vn_F$ as
a predefined normal over each $F$ being coincident with $\Vn$ when ${F \in \Sk_h^\partial}$, and $h_F$ as the diameter of the face $F$. On interior faces, for any function ${v \in V_{h, \#}}$, the jump  $[\![v]\!]_F$ and the standard (arithmetic) average $\{v\}_F$ are defined as 
\begin{align*}
[\![v]\!]_F := v^- - v^+\quad \quad , \quad \quad \{v\}_F:= \frac{1}{2}(v^- + v^+),
\end{align*}
with $v^-$ and $v^+$ denoting the left and right face values respectively, with respect to the predefined normal $\Vn_F$. For ${F \in \Sk^\partial_h}$, we set ${[\![v]\!]_F=\{v\}_F=v|_F}$. 

Finally, for a given norm $(\boldsymbol{\cdot},\boldsymbol{\cdot})_{V_h}$ of the discrete space $V_h$, we define the dual norm $\| \boldsymbol{\cdot} \|_{V^*_h}$ for $\phi \in V^*_h$ as:
\begin{align}\label{eq:dual-norm}
\| \phi \|_{V^*_h} := \sup_{v_h \in V_h \backslash \{0\}} \frac{\langle \phi, v_h \rangle_{V^*_h \times V_h}}{\| v_h \|_{V_h}},
\end{align}
where $\langle \boldsymbol{\cdot}, \boldsymbol{\cdot} \rangle_{V_h^* \times V_h}$ denotes the duality pairing in $V_h^* \times V_h$.

\subsubsection{Nonlinear discontinuous Galerkin formulation}
Considering the above discrete setting, we build the dG formulation for the continuous weak variational formulation  of~\eqref{eq:weak_nonlin} as
\begin{equation}\label{eq:dg_nonlin}
\left\{\begin{array}{l}
\text{Find } u_h \in V_h, \text{ such that:} \smallskip \\
\begin{array}{l}
n_h(u_h; v_h)= \ell_h(v_h), \quad \forall v_h \in V_h,
\end{array}
\end{array}\right.
\end{equation}
with
\begin{equation*}
\begin{array}{rl}
\displaystyle n_h(u_h;v_h) := & \displaystyle \sum_{T \in \Par_h} (\nabla u_h \, , \, \nabla v_h)_{T} + \sum_{T \in \Par_h} (\mathcal{F}(u_h)\, , \,v_h)_{T} \\
& + \displaystyle \sum_{F \in \Sk_h}\left[ \left( [\![ u_h ]\!] \, , \,  \{  \nabla v_h \} \cdot \Vn_F \right)_F - \left( \{  \nabla u_h \} \cdot \Vn_F  \, , \,  [\![ v_h ]\!]\right)_F +  \dfrac{\gamma}{h_F}\left( [\![ u_h ]\!], [\![ v_h ]\!] \right)_F \right], \\
\displaystyle \ell_h(v_h) :=& \displaystyle \sum_{T\in\Par_h}(f, v_h)_T.
\end{array}
\end{equation*}
In the above, $\gamma>0$ is a user defined constant that we set as $\gamma = 3(k+1)(k+2)$, being $k$ the polynomial degree of the test space. Besides, we recall~\eqref{eq:der-def} and set the discrete derivative as:
\begin{equation}\label{eq:cnoidal_linearized} 
\begin{array}{rcl}
\displaystyle n'_h(u_h; z_h, v_h)& :=&\displaystyle\frac{d}{d\epsilon}n_h(u_h+\epsilon z_h; v_h)\big|_{\epsilon=0} \vspace{0.2cm}\\
&=&\displaystyle \sum_{T \in \Par_h} (\nabla z_h \, , \, \nabla v_h)_{T} + \sum_{T \in \Par_h} (\mathcal{F}'(u_h) \, z_h\, , \,v_h)_{T} \\
& +&\displaystyle \displaystyle \sum_{F \in \Sk_h}\left[ \left( [\![ z_h ]\!] \, , \,  \{  \nabla v_h \} \cdot \Vn_F \right)_F - \left( \{  \nabla z_h \} \cdot \Vn_F  \, , \,  [\![ v_h ]\!]\right)_F +  \dfrac{\gamma}{h_F}\left( [\![ z_h ]\!], [\![ v_h ]\!] \right)_F \right],
\end{array}
\end{equation}
where 
\begin{equation}
\mathcal{F}'(u_h)=\lambda \, m \, (1+u_h)^{m-1} - \mu \, \beta \exp(\beta u_h).
\end{equation}
We build our resolution scheme using~\eqref{eq:cnoidal_linearized}. This linearized form can be seen as a reaction-diffusion form in each increment $z_h$, for that reason, we provide the discrete space $V_h$ with a diffusion-type norm: 
\begin{align}\label{eq:diff-norm}
\|w\|_{V_h}^2 :=& \; \theta \ \|w\|^2_{0, \Omega} + \| \nabla w\|^2_{0, \Omega} +  \sum_{F \in \mathscr{S}_h} \left(  \dfrac{\gamma}{h_F}[\![w]\!] , [\![w]\!] \right) _{0,F}.
\end{align}
with ${\theta=\lambda \, m \, A^{m-1}}$, where ${A>0}$ is a given constant associated to the maximum value of the normalized stress in the cnoidal solution. 

\subsection{Adaptive stabilized finite element method based on residual minimization}
The stabilized finite element method based on residual minimization presented in~\cite{calo2019} delivers a mixed problem, with a saddle-point structure, in the case of linear problems. In this section, we develop the discrete formulation for the continuous problem~\eqref{eq:abs_vf}, which reads: 
\begin{equation}\label{eq:nonlin}
{N_h(u_h)=\ell_h(\boldsymbol{\cdot})},
\end{equation}
where ${N_h : U_h \rightarrow V_h^*}$ represents the discrete nonlinear map with ${\langle N_h(z_h), v_h\rangle_{V_h^* \times V_h} := n_h(z_h; v_h)}$, being $n_h(\boldsymbol{\cdot}\, ;\boldsymbol{\cdot})$ the nonlinear form associated with problem~\eqref{eq:dg_nonlin}.

The methods solves for a \textit{continuous} approximation solution in a given discrete space $U_h$ (for instance, standard FEM functions), adequately considered as a subspace of the dG space $V_h$, i.e., ${u_h \in U_h = V_h \cap C^0(\Omega) \subset V_h}$. For example, in~\eqref{eq:dg_nonlin}, ${u_h \in U_h = V_h \cap H^1(\Omega)}$. The solution $u_h$ is then computed through minimizing the residual ${\ell_h(\cdot) - N_h (z_h)}$ associated to~\eqref{eq:nonlin}  in the norm of $V_h^*$:
\begin{align}\label{resmin_nonlin}
u_h = \operatorname*{argmin}_{z_h \in U_h} \frac{1}{2} \|\ell_h(\boldsymbol{\cdot}) - N_h (z_h)\|^2_{V_h^*} = \operatorname*{argmin}_{z_h \in U_h}\frac{1}{2} \|R_{V_h}^{-1} (\ell_h(\boldsymbol{\cdot})- N_h (z_h)) \|^2_{V_h},
\end{align}
with $\| \boldsymbol{\cdot} \|_{V_h}$ the norm defined in~\eqref{eq:diff-norm}, and $R^{-1}_{V_h}$ the inverse of the Riesz map (cf.,~\cite{oden2017}, Theorem 6.4.1). Similar to the thinking for~\cite{calo2019}, the nonlinear problem can be stated as a critical point of the minimizing functional, which translates into the following linear problem: Find $u_h \in U_h$ such that:
\begin{align}\label{eq:resmin_as_nonLinProb}
(R_{V_h}^{-1}(\ell_h - N_h(u_h)),R_{V_h}^{-1}DN_h(u_h; z_h))_{V_h} = 0, \quad \forall z_h \in U_h,
\end{align} 
being $DN_h(u_h; z_h)$ the discrete form of the derivative (cf.,~\eqref{eq:der-def}). As noticed in~\cite{cohen2012}, problem~\eqref{eq:resmin_as_nonLinProb} can be equivalently written as the following saddle-point problem:

\begin{equation}\label{eq:nl-femwdg}
\left\{\begin{array}{l}
\text{Find } (\varepsilon_h, u_h) \in V_h \times U_h,  \text{ such that:} \smallskip \\
\begin{array}{rcll}
(\varepsilon_h, v_h)_{V_h} + n_h(u_h;v_h) &=& \ell_h(v_h), & \quad  \forall v_h \in V_h, \smallskip \\
n'_h(u_h; z_h, \varepsilon_h) &=& 0, &\quad  \forall z_h \in U_h.
\end{array}
\end{array}\right.
\end{equation}
The system~\eqref{eq:nl-femwdg} delivers simultaneously a stable and continuous approximation $u_h \in U_h$ of the dG formulation, and a residual representation $\varepsilon_h \in V_h$ that guides the adaptive mesh refinement. The first line of this system represents the nonlinear problem to solve, associated to the residual, whereas the second line can be seen as a constraint imposed on the tangent space built from the linearized form.

\subsubsection{Linearized problem}

We use Newton's method for solving the nonlinear problem. Given the discrete solution pair ${(\varepsilon^i_h, u^i_h)}$ of an iterative step $i$, we look for the increment ${(\delta \varepsilon_h, \delta u_h)}$ of the next iteration, and we set ${u}^{i+1}_h={u}^{i}_h+t^i {\delta u}_h$, and ${\varepsilon}^{i+1}_h={\varepsilon}^{i}_h+t^i {\delta \varepsilon}_h$, where $t^i$ represents a relaxation parameter to control the increment size. The method seeks for the solution pair $(\varepsilon_h^{i+1}, u_h^{i+1})$ that satisfies~\eqref{eq:nl-femwdg}. For the $i+1$-th iteration, the linearization of~\eqref{eq:nl-femwdg} reads:
\begin{equation}\label{eq:nl-newton}
\left\{\begin{array}{l}
\text{Given the pair } (\varepsilon^i_h, u^i_h), \text{ find } (\delta \varepsilon_h, \delta u_h) \in V_h \times U_h,  \text{ such that:} \smallskip \\
\begin{array}{rcll}
\hspace{-0.25cm}(\delta \varepsilon_h, v_h)_{V_h} + n'_h(u^i_h; \delta u_h, v_h) & \hspace{-0.25cm}= & \hspace{-0.25cm} \ell_h(v_h)- (\varepsilon^i_h, v_h)_{V_h} -n_h(u^i_h; v_h) , & \quad \forall v_h\in V_h,   \\
\hspace{-0.2cm}n'_h(u^i_h; z_h, \delta \varepsilon_h) & \hspace{-0.25cm}= & \hspace{-0.25cm}-n'_h(u^i_h; z_h, \varepsilon^i_h),                              & \quad \forall z_h \in U_h.
\end{array}
\end{array}\right.
\end{equation}

\begin{figure}[h!]
        \centering
        \includegraphics[width=0.8\textwidth]{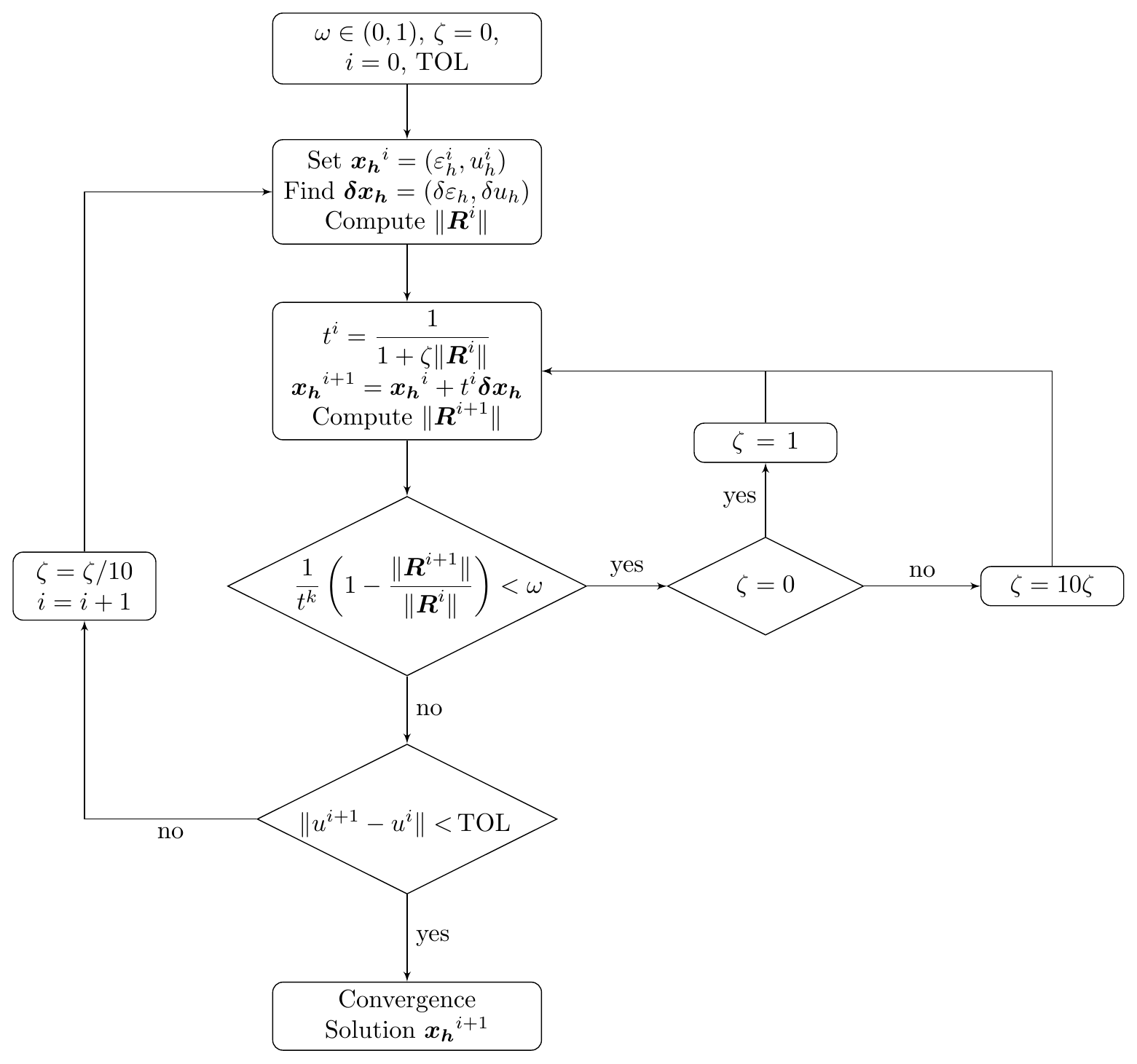} 
        \caption{Damped Newton's algorithm flow chart.}
        \label{fig:flow-chart}
\end{figure}

In matrix form, formulation~\eqref{eq:nl-newton} reads:
\begin{align}\label{eq:femwdg_mtx}
\begin{pmatrix}
\ G & B_u \ \\ \ B_u^T & 0 \ 
\end{pmatrix}
\begin{pmatrix}
\ \delta \varepsilon_h \ \\
\ \delta u_h \
\end{pmatrix}
=
\begin{pmatrix}
\ L  \ \\
\ 0 \
\end{pmatrix}
-
\begin{pmatrix}
\ G \varepsilon^i_h + N(u^i_h)\, \ \\
\ B^T_u \varepsilon^i_h \,\
\end{pmatrix}
\end{align}
where $G$ is the Grammian matrix built for the inner product that induces the norm in the discrete space $V_h$, $N(u^i)$ is the vector associated to the nonlinear form $n_h(u_h;v_h)$ and $B_u$ is the matrix associated with its linearization $n'_h(u_h^{i}; \delta u_h, v_h)$. The residual representative $\varepsilon_h$ is an implicit function of $u_h$. We define the pair ${\boldsymbol{ x_h}=( \varepsilon_h,  u_h)}$ that comprises both the solution and the residual representative, being valid also for the increments, which allows us to rewrite~\eqref{eq:femwdg_mtx} as:
\begin{equation*}
\boldsymbol{J}^i \, \boldsymbol{\delta x_h} = \boldsymbol{R}^i,
\end{equation*}
where 
\begin{equation*}
\begin{array}{c}
\boldsymbol{J}^i =
\begin{pmatrix}
\ G & B_u \ \\ \ B_u^T & 0 \ 
\end{pmatrix}
\, \; \text{ and } \; \,
\boldsymbol{R}^i
=
\begin{pmatrix}
\ L  \ \\
\ 0 \
\end{pmatrix}
-
\begin{pmatrix}
\ G \varepsilon^i_h + N(u^i_h)\, \ \\
\ B^T_u \varepsilon^i_h \,\
\end{pmatrix}
\end{array}
\end{equation*}
The convergence of the method is controlled by the size of each iteration step through the relaxation parameter $t^i$. For that purpose, we use the damped Newton's method~\cite{bank1981} shown in Figure~\ref{fig:flow-chart}.

\section{One-dimensional numerical experiments}\label{sec:numexp_1d}

In this section, we develop several one-dimensional numerical examples to illustrate the performance of the adaptive stabilized finite element method in the context of the cnoidal equation. Simulation results reported use FEniCS~\cite{ Alnaes2015} in \S \ref{ss:single-peak} \& \S \ref{ss:mult-peak}, and REDBACK~\cite{Tung2017} in \S \ref{ss:cont}. The main drawback of standard FEM implementations for this kind of problem lies in the difficulty to resolve the localized peak and particularly to find their location, which ends up delivering low-quality solutions. These limitations severely restrict its usage, forcing the initial guess to be close enough to the actual solution for the algorithm to converge, which is impractical. To overcome this these limitations in a nonlinear framework, we seek an algorithm with the ability to find automatically the location of the peaks. In practice, this means that we can start from an arbitrary initial trial solution with peaks located far from the final configuration. The numerical examples demonstrate that the adaptive stabilized finite element method (cf., \S\ref{sec:num}) can easily overcome these issues. 

\subsection{Single peak solution}\label{ss:single-peak}

\begin{figure}[h!]
    \centering
    \begin{subfigure}[c]{0.49\textwidth}
        \centering
        \includegraphics[width=\textwidth]{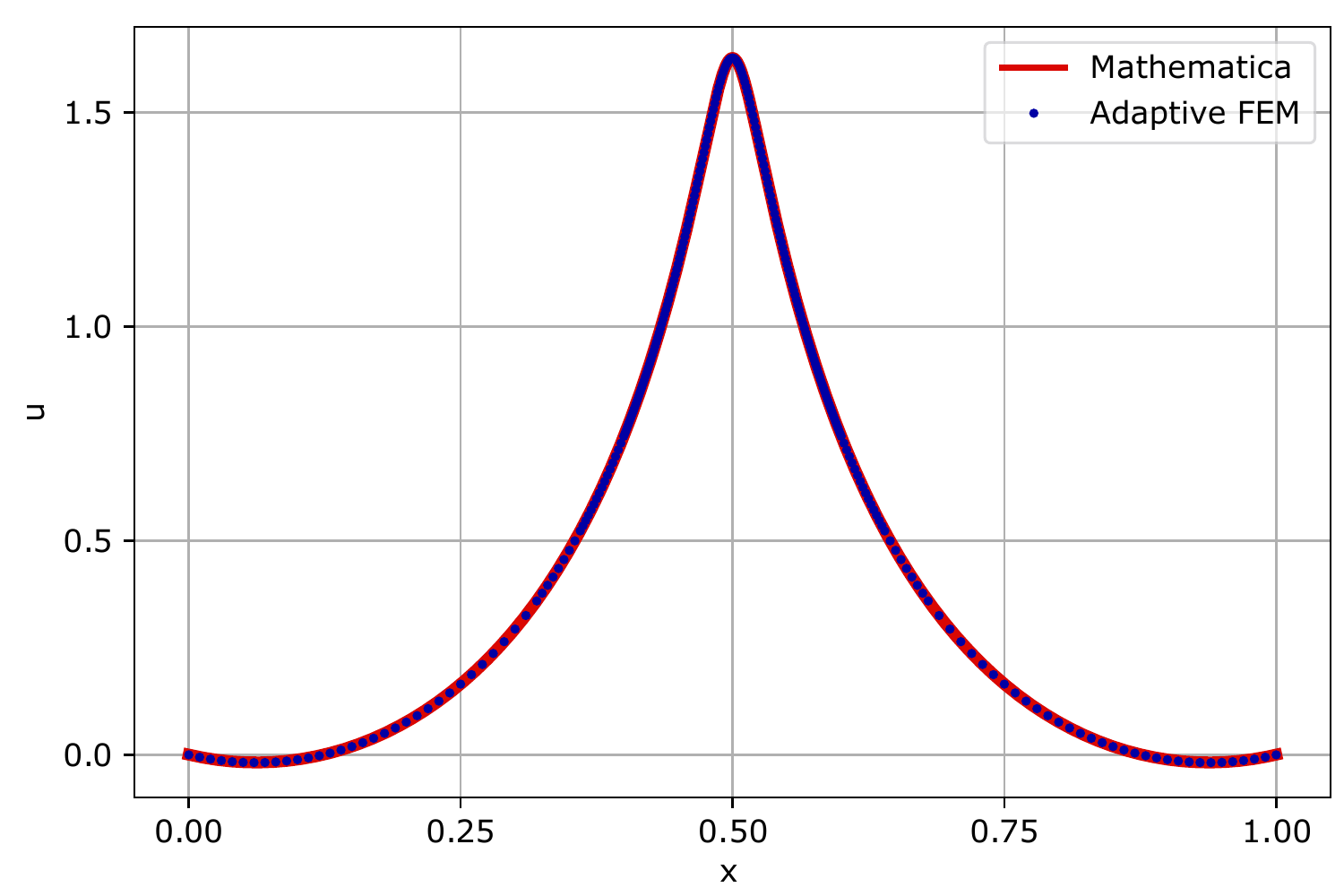} 
        \caption{\centering{Whole solution.}}
        \label{fig:1peak}
    \end{subfigure}
    \begin{subfigure}[c]{0.49\textwidth}
        \centering
        \includegraphics[width=\textwidth]{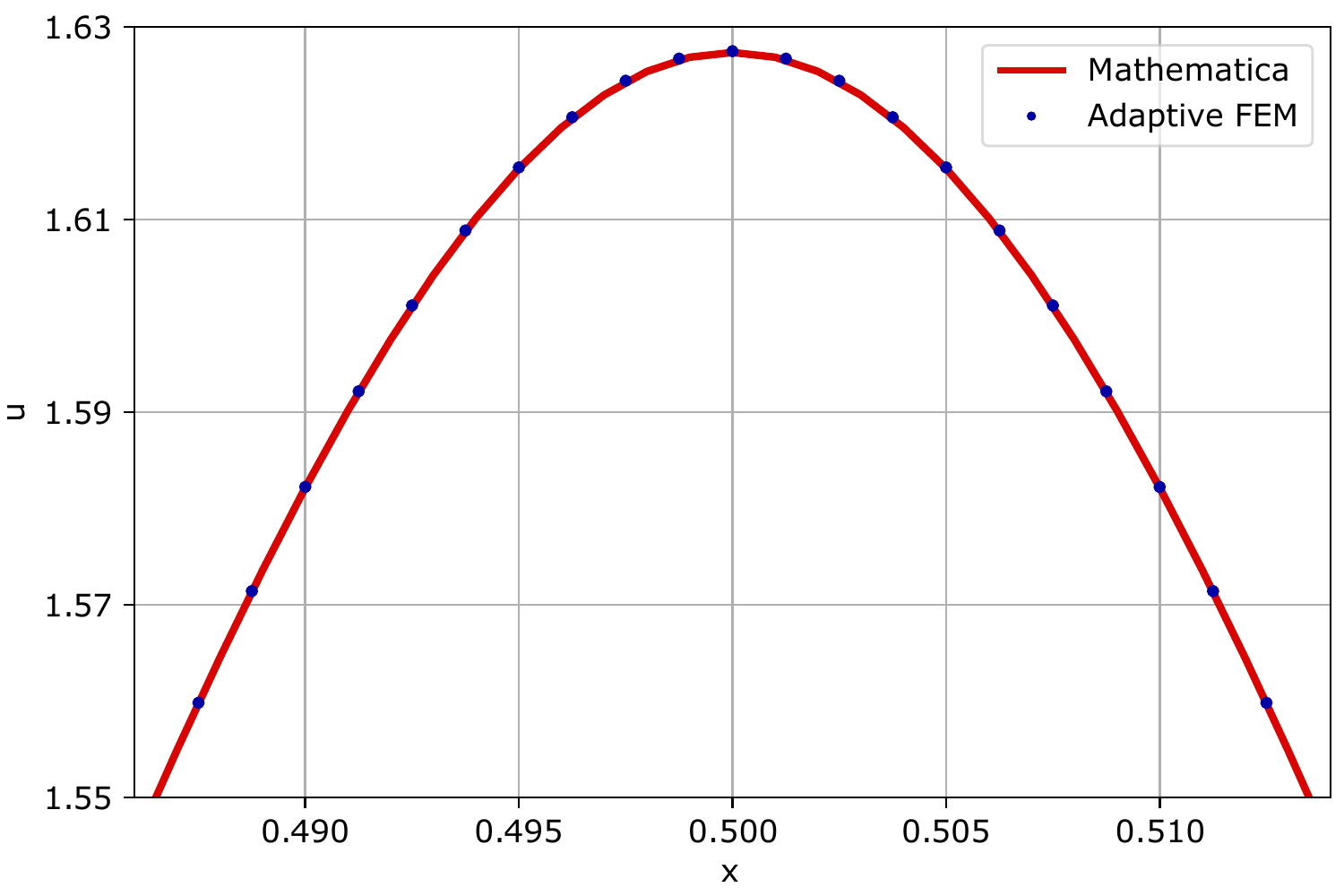}
        \caption{\centering{Zoom around $x=0.5$.}}
        \label{fig:1peak-zoom}
    \end{subfigure}
    \caption{Comparison showing the good match between the Mathematica solution and our results for $\lambda=10$, starting from the initial guess ${u_{IG}=2 \exp(-100(x-0.5)^2)}$.}
    \label{fig:comparison_mathematica}
\end{figure}

Thanks to the enhanced stability of the framework, we can solve~\eqref{eq:numsol_def_equation} and retrieve the expected peak solution for appropriate parameters. Figure~\ref{fig:comparison_mathematica} shows the comparison between the semi-analytical solution computed with Mathematica~\cite{wolfram1999mathematica} and the results obtained with our approach for ${\lambda=10,\, m=3, \,\mu=10^{-4}}$ and ${\beta=10}$, starting from an initial guess ${u_{IG}=2 \exp({-100(x-0.5)^2})}$ on a regular mesh of 100 nodes, getting to 273 nodes after four levels of adaptivity. We observe an excellent match, including the peak location, shape, and intensity, as shown in Figure~\ref{fig:1peak-zoom}.

\subsection{Multiple peak solution}\label{ss:mult-peak}
Given that we can recover known semi-analytic solutions with the method, we explore a harder problem considering more peaks. In this context, standard FEM approach is no longer appropriate, due to its lack of stability. For the following numerical examples, we define 
\begin{equation}\label{eq:u_ig_1d}
{u_{IG}(x) := A_0 \, \left[\frac{\exp({-1250(x-x_0)^2})}{\sin(x_0\pi)}+ \frac{\exp({-1250(x-(1-x_0))^2})}{\sin((1-x_0)\pi)}\right]\sin(x\pi) }
\end{equation}
as the initial guess function, being $x_0$ the arbitrary location of the first peak ${(0 \leq x_0\leq 0.5)}$. This initial guess choice implies that the second peak location is at $1-x_0$. Besides, $A_0>0$ is an arbitrary number that coincides with the values of $u$ at the peak locations in the initial guess. Table~\ref{tab:param} shows the set parameters for the examples in this subsection.

\begin{table}[ht!]
	\small
	\centering
	\caption{Parameters for the one-dimensional numerical examples with two-peak solution.}
	\begin{tabular}{|c|c|c|c|c|c|c|} 
		\hline
	 	Example&$\lambda$ &$m$ & $\mu$& $\beta$ & $A_0$ & $x_0$ \\
	 	\hline
		3.2.1&40&3&$10^{-4}$&10&1.80&0.200\\
		\hline
		3.2.2&40&3&$10^{-4}$&10&1.80&0.175\\
		\hline
		3.2.3&40&3&$10^{-4}$&10&1.80&0.425\\
		\hline
		3.2.4&40&$\pi$&$10^{-4}$&10&2.70&0.350\\
		\hline
	\end{tabular}
	\label{tab:param}
\end{table}

\subsubsection{Comparison against standard FEM}
\begin{figure}[h!]
	\centering
	\includegraphics[width=0.6\textwidth]{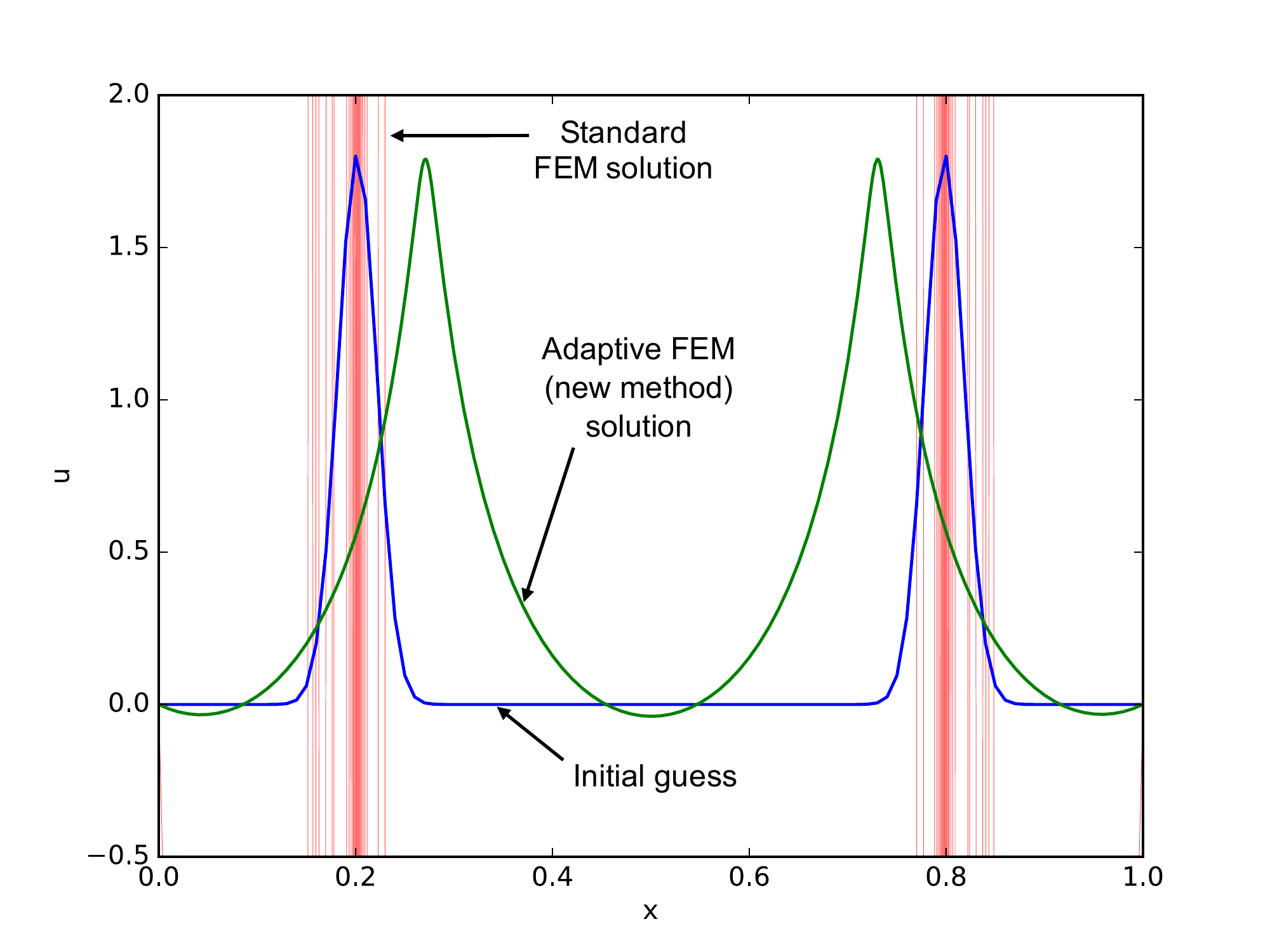}
	\caption{Comparison between standard finite element solution (in red) and the adaptive stabilized method (in green) for an initial guess ${u_{IG} := 1.8 \, [\exp({-1250(x-0.2)^2})/\sin(0.2\pi)+ \exp({-1250(x-0.8)^2})/\sin(0.8\pi)]\sin(x\pi)}$ (in blue). The stabilized method converges to a stable solution, whereas the standard FEM approach leads to spurious oscillations.}
	\label{fig:sol-2peaks}
\end{figure}

As a first example, Figure~\ref{fig:sol-2peaks} shows the results comparing the discrete solution obtained with the standard FEM formulation and the new adaptive stabilized method, for the same initial guess. For this two-peak example, we set the arbitrary location as ${x_0=0.2}$ (see Table~\ref{tab:param}). We use cubic trial functions $(\mathds{P}_3)$ for both methods, but we take advantage of the possibility of enriching the test space in the adaptive stabilized method, using test functions one degree higher $(\mathds{P}_4)$. Finally, we use a fixed mesh for the standard finite element solution of mesh size of ${h = 10^{-6}}$, to develop a fair comparison with the final refined mesh obtained through the adaptive method, which starts from a mesh size of 100 elements (${h=0.01}$) and gets no finer than ${h=10^{-6}}$ locally. We can observe that this new technique properly captures the final location of the peaks at $x\approx\{0.27, 0.73\}$, whereas the standard method delivers spurious oscillations. 

\subsubsection{Initial guess close to the boundaries}
\begin{figure}[h!]
	\centering
	\includegraphics[width=0.6\textwidth]{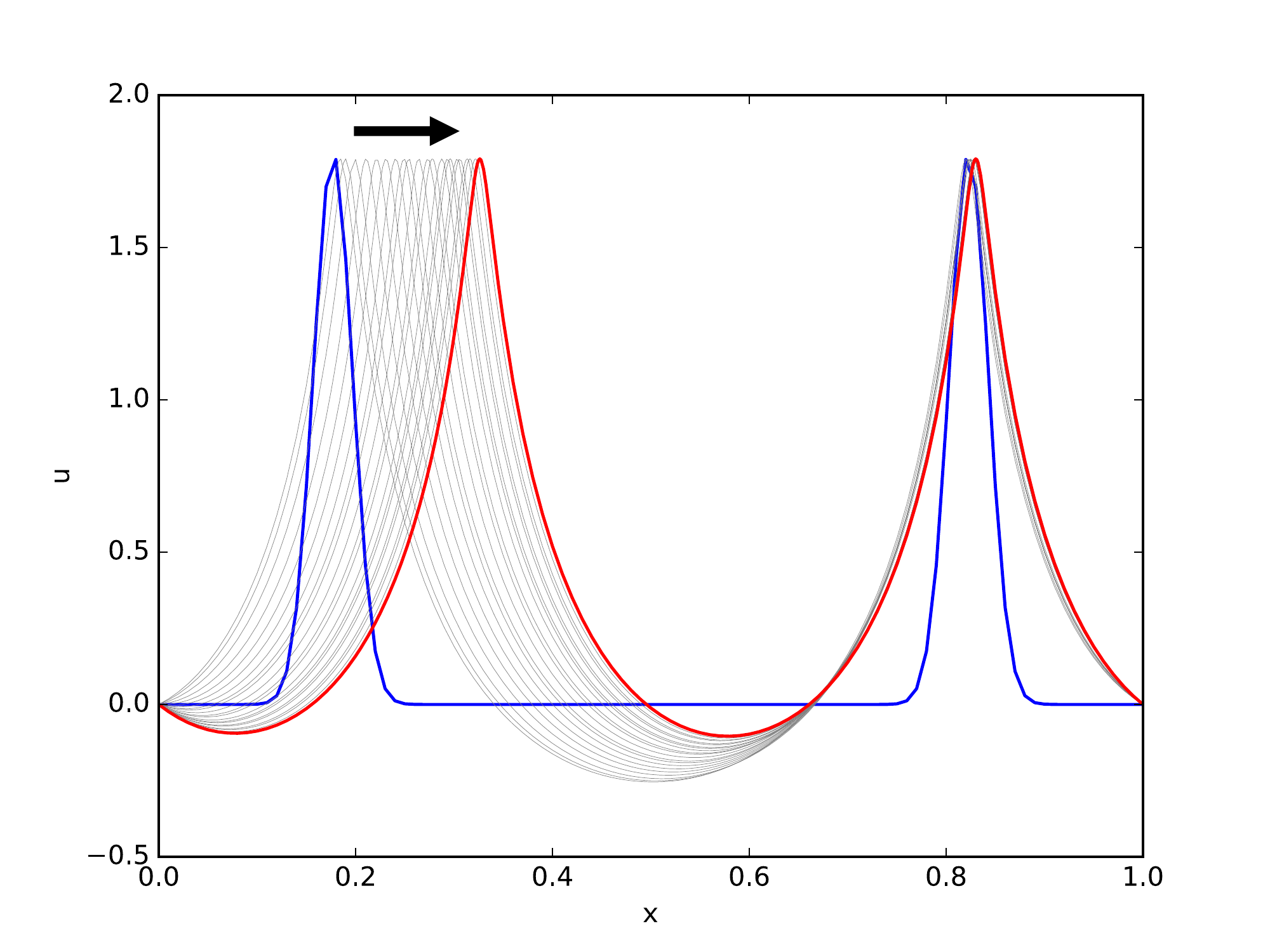} 
	\caption{Evolution of the profiles at intermediate mesh refinement steps for an initial guess using ${x_0=0.175}$ (in blue) showing convergence to an appropriate (asymmetrical) solution (in red).}
	\label{fig:2peaks-175}
\end{figure}

We now investigate examples with different conditions to show that the new method converges robustly with respect to the initial condition. The distance  between the initial and final peak locations using the adaptive method is significantly larger than that of the one for which standard FEM on a fine mesh can converge. In this example, we locate the peaks close to the boundaries (see Table~\ref{tab:param}). Figure~\ref{fig:2peaks-175} shows the iterative solutions obtained at each refinement step. The adaptive method converges when standard FEM fails, even with an order of magnitude finer mesh (${h=10^{-7}}$). The adaptive approach successively corrects the peak locations at each refinement level, starting from a mesh size of $h=0.01$. After 47 refinement levels (approximately, $24,000$ iterations), we obtain a solution with a final residual norm $\|\boldsymbol{R}^{i+1}\|_0<10^{-9}$. Incidentally, this example also shows that the solution can be asymmetric.

\subsubsection{Initial guess close to the center}
\begin{figure}[h!]
	\centering
	\includegraphics[width=0.6\textwidth]{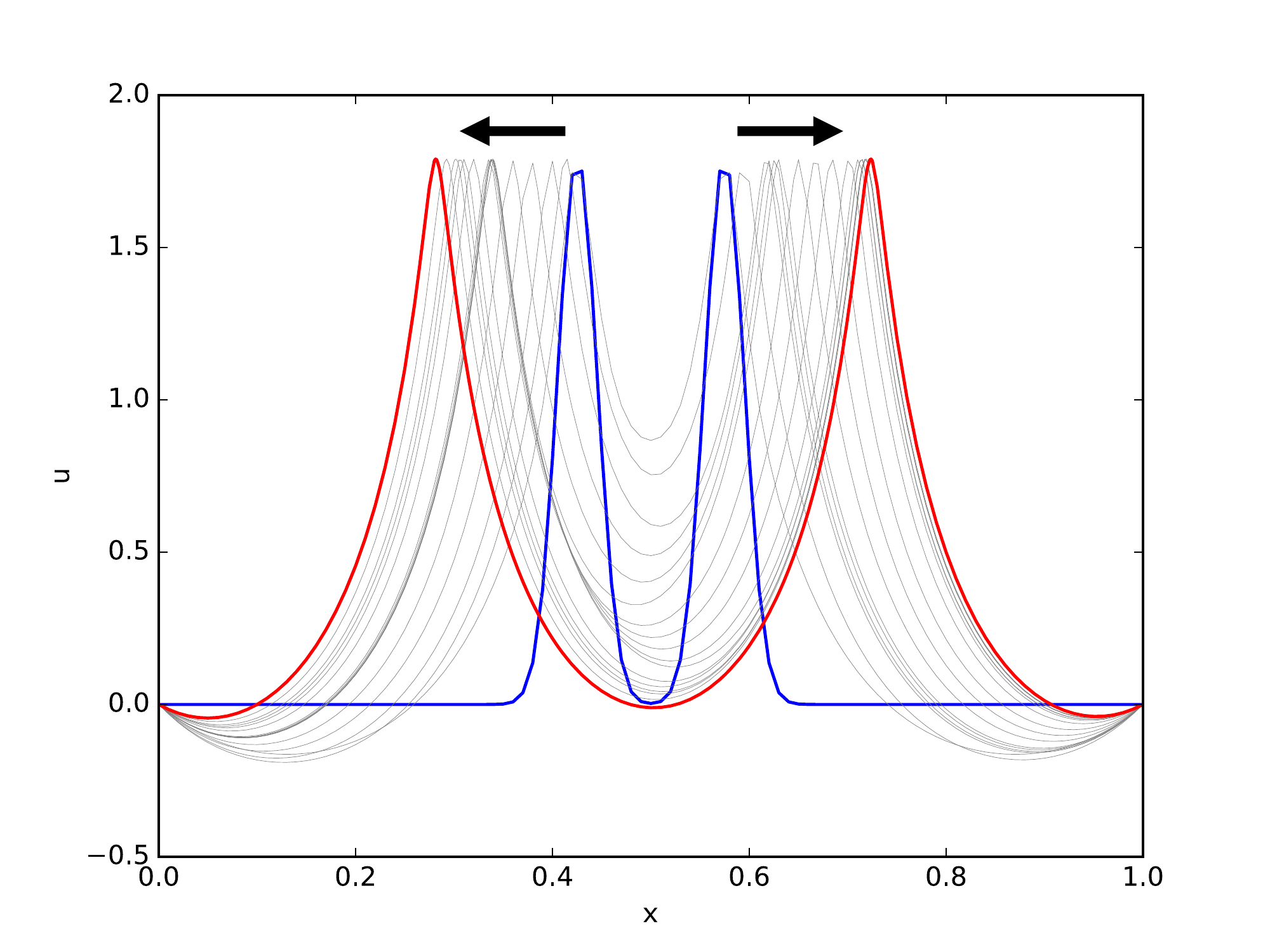} 
	\caption{Evolution of the profiles at intermediate mesh refinement steps for an initial guess using ${x_0=0.425}$ (in blue), showing convergence to an appropriate (symmetrical) solution (in red).}
	\label{fig:2peaks-425}
\end{figure}

In this example, we locate the initial peaks close to the center (see Table~\ref{tab:param}). In that instance, we end up converging to the solution of Figure~\ref{fig:2peaks-425} using the adaptive stabilized method after 37 refinement levels (approximately, $7,400$ iterations) for the same tolerance than in the previous case. The final peak locations are $x\approx\{0.27, 0.73\}$. FEM is not able to converge using this initial guess either. From our experience, which we do not report for the sake of brevity, we find that FEM simulations require initial guesses sufficiently close to the final solution for the method to converge. In practice, FEM requires the distance between the initial and the final solution peak to be at least an order of magnitude smaller than what the adaptive stabilized method admits.

\subsubsection{Non-integer exponent}
\begin{figure}[h!]
    \centering
    \includegraphics[width=0.6\textwidth]{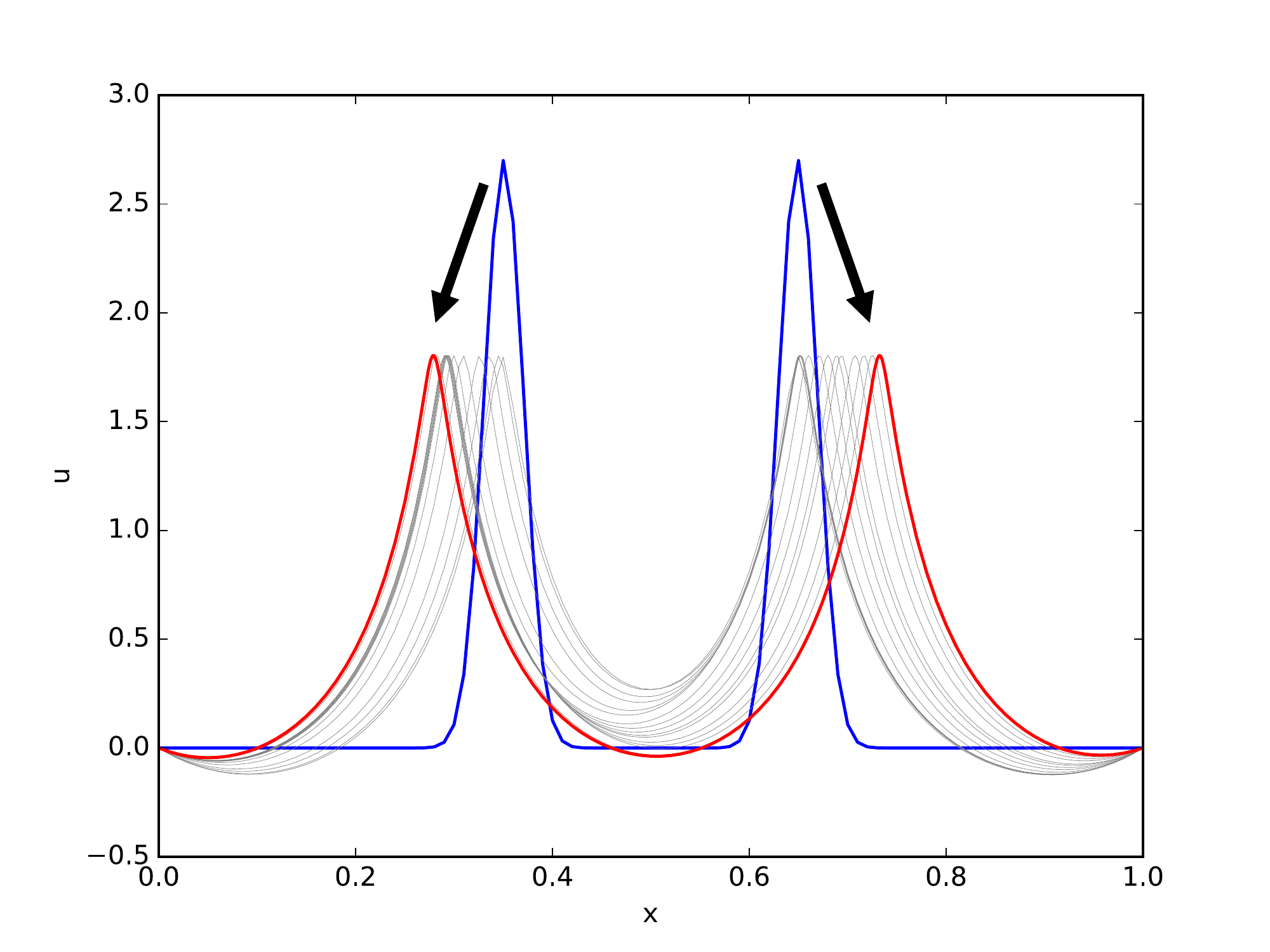}
    \caption{Evolution of the two-peak solution profile using the adaptive stabilized method for $m=\pi$ and an initial guess with peak values of ${u(0.35)=u(0.65)=2.7}$ (in blue), showing convergence to an appropriate (symmetrical) solution (in red).}
    \label{fig:sol-pi}
\end{figure}

Finally, we simulate a scenario with a non-integer exponent $m=\pi$ to show the robustness of the adaptive stabilized method for an irrational exponent. As mentioned in \S\ref{sec:num}, the analytical approach of the cnoidal equation does not provide solutions for this class of exponents. We consider the initial guess~\eqref{eq:u_ig_1d} with $x_0=0.35$ and we set higher peaks values than in past examples (see Table~\ref{tab:param}). Figure~\ref{fig:sol-pi} displays the evolution of the solution profile, showing that the method can robustly simulate irrational exponents larger than 3, a limitation of the analytical resolution approach~\cite{Veveakis2015}. Also, the adaptive stabilized scheme correct the height of the peak values. Numerical simulations not presented here also showed a good performance for even higher exponent values up to 7.

\subsection{Advantages of the usage of the adaptive FEM method in the cnoidal problem}
In summary, the adaptive stabilized method can robustly and efficiently predict the solution patterns for many relevant configurations for the cnoidal wave problem. This heretofore unsurmountable problem is now solvable and the performance of the method can be explained for two reasons. Firstly, this technique allows the enrichment of the test space, which implies having stronger approximation norms and better behavior in the solution at each level. This enrichment is not possible for FEM, due to the structure of Galerkin's approach. Secondly, the adaptive mesh refinement scheme ensures the elimination of oscillations through the reduction of the local error. This robust adaptivity represents an important feature of the method, especially because of the localized nature of the solution. Using this adaptive stabilized method, we can build an indicator from the residual representative that allows us to correct the location of the peaks at each refinement level.

\subsection{Numerical continuation}\label{ss:cont}

We can now investigate in detail the influence of the parameter $\lambda$, to better understand the most notable aspects of the solution. In particular, we are interested in identifying the range of $\lambda$ values where a peak solution exists. More precisely, we seek to discern the presence of a threshold $\lambda_c$, above which multiple peaks can be obtained, as well as the relationship between the number, $p$, of peaks and $\lambda$. For example,~\cite{Regenauer-Lieb2013a} reported a ${p\propto\sqrt{\lambda}}$. To that effect, we use continuation technique to study the stability regimes of the system, starting from identified solutions.

We note that it is possible to solve the cnoidal equation~\eqref{eq:cnoidal_1} in the stationary limit via standard FEM procedures starting from a sufficiently accurate initial solution, which allows us to retrieve the continuation map using classical techniques. Hence, we perform stability analysis using a pseudo-arclength continuation algorithm from~\cite{Keller1977}, which allows us to identify different behavior characteristics of the solution of the system (\ref{eq:cnoidal_chem}) where we also consider $\lambda$ as a variable. The algorithm starts with two solutions $(u_{0}, u_1)$ of (\ref{eq:cnoidal_chem}) for the respective values $(\lambda_0, \lambda_1)$, where each solution is computed on a discretized mesh of $n$ points. The next step is then computed starting from the initial guess ${[2u_1-u_{0},2\lambda_1-\lambda_0]^T}$, i.e. following the tangent vector of $[u,\lambda]^T$ in $\mathbb{R}^{n+1}$ and searching for a solution $[u,\lambda]^T$ at a distance $\Delta s$ from the previous one. We compute following steps recursively. The extended system of equations solved can then be written as
\begin{subequations}
    \label{eq:continuation_system}
    \begin{align}
    & \frac{\partial^2 u_n}{\partial z^2}-\lambda(1+u_n)^{m}+\mu \mathrm{e}^{\beta u_n}=0, \\
    &\dot{u}_{0}\left(u-u_{0}\right)+\dot{\lambda}_{0}\left(\lambda-\lambda_{0}\right)=\Delta s,
    \end{align}
\end{subequations}
where $[\dot{u}_{0},\dot{\lambda}_{0}]^T$ is the tangent vector at $[u_{0},\lambda_{0}]^T$.

\subsubsection{One peak solutions}

\begin{figure}[t!]
    \centering
    \begin{subfigure}[c]{0.8\textwidth}
        \centering
        \includegraphics[width=\textwidth]{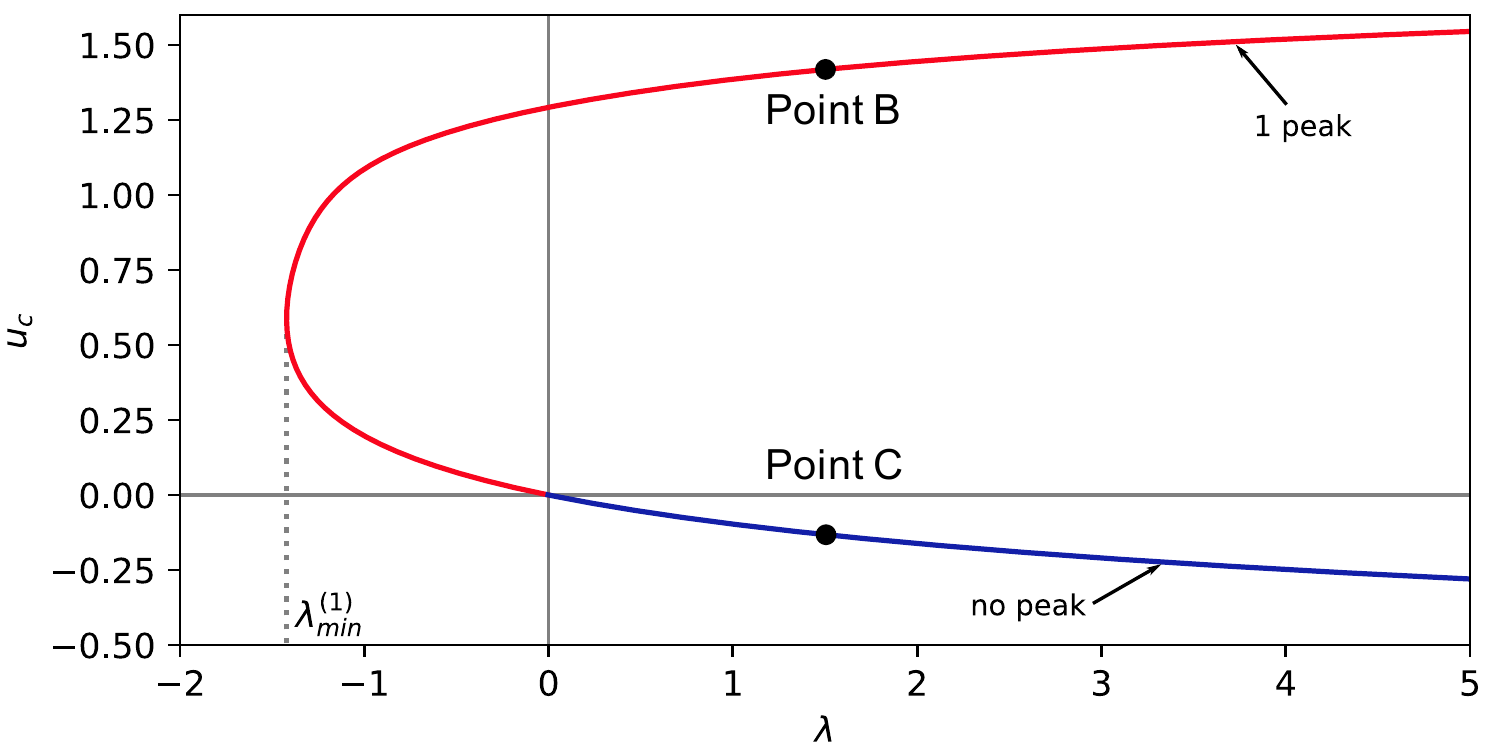}
        \caption{} 
        \label{fig:c-curve-plot}
    \end{subfigure}
    \begin{subfigure}[c]{0.4\textwidth}
        \vspace{0.5cm}
        \centering
        \includegraphics[width=\textwidth]{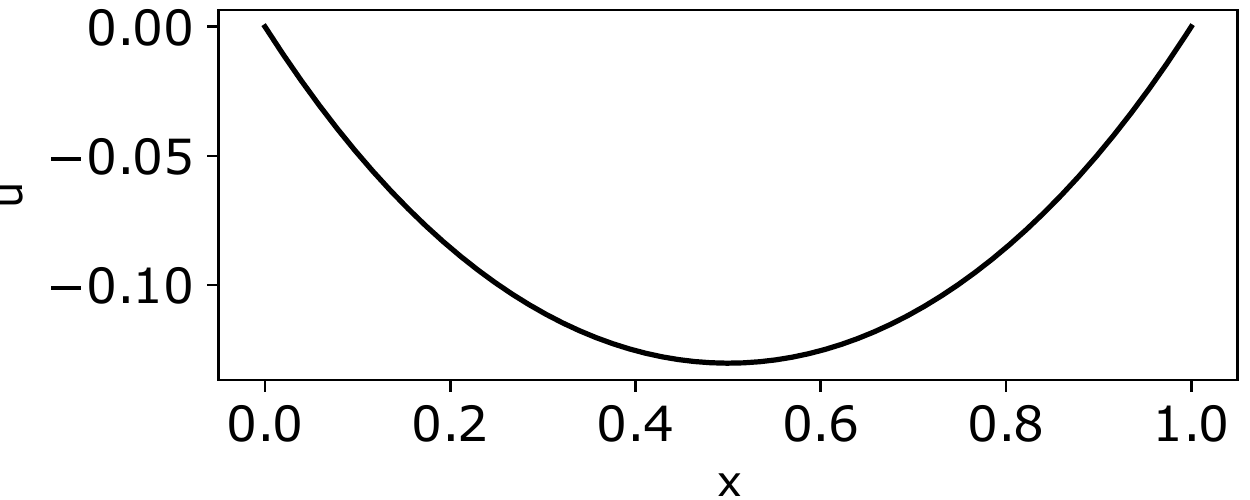} 
        \caption{} 
        \label{fig:ccurve-point-b}
    \end{subfigure}
    \begin{subfigure}[c]{0.4\textwidth}
        \vspace{0.5cm}
        \centering
        \includegraphics[width=\textwidth]{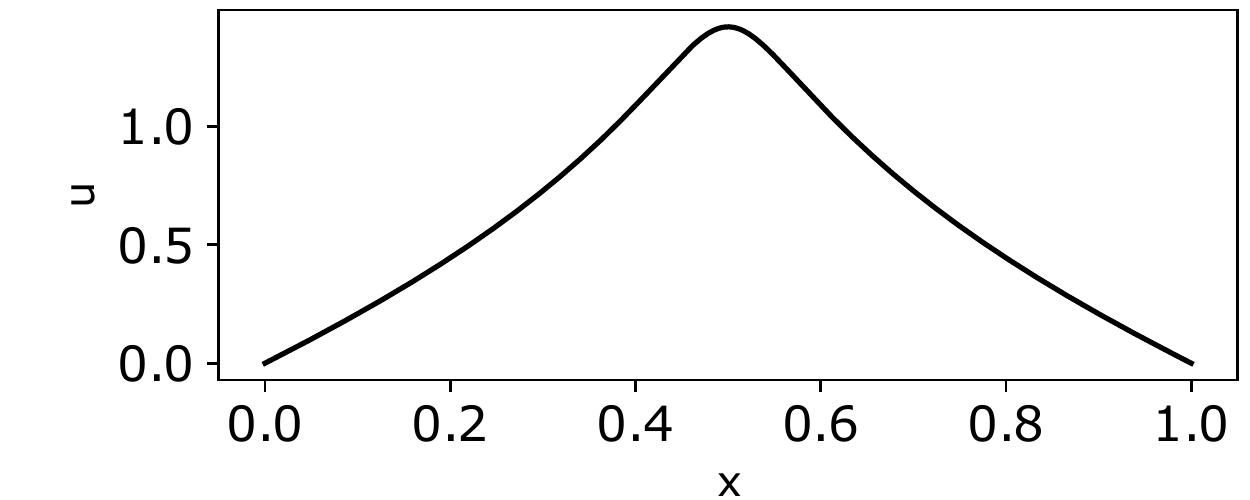}
        \caption{} 
        \label{fig:ccurve-point-c}
    \end{subfigure}
    \caption{(a) C-curve result of the stability analysis of (\ref{eq:cnoidal_chem}), plotting the value $u_c$ of the solution at the center of the domain ($x=0.5$) with respect to parameter $\lambda$, capturing the traditional Terzaghi's consolidation (blue lower branch) and single peak case (red upper branch). Points B and C mark two points on the lower and upper branches for the same value of $\lambda=1.5$ and their respective solution profiles are shown in subfigures (b) and (c); (b) Terzaghi's solution profile for $\lambda=1.5$; (c) one-peak solution profile for $\lambda=1.5$}
    \label{fig:C-curve-1peak}
\end{figure}

Solving (\ref{eq:cnoidal_chem}) for any positive value of $\lambda$ with a zero initial guess leads to the traditional Terzaghi's consolidation isochrone profile $u(x)\leq0,  \: \forall x \in [0,1]$~\cite{Terzaghi1943} shown in Figure~\ref{fig:ccurve-point-b}. Running the numerical continuation algorithm from two such initial solutions for increasing values of $\lambda$ yields the expected branch of traditional Terzaghi's solutions, with the minimum value $u_c$, obtained at the center ($x=0.5$) for symmetry reasons, decreasing asymptotically towards $-1$ for ${\lambda \to \infty}$. Running the continuation algorithm for decreasing values of $\lambda$ provides the evolution of the solution as $\lambda$ becomes negative. Figure~\ref{fig:c-curve-plot} shows the results plotted in the ${\lambda-u_c}$ space. We observe a C-shape solution, marked by a threshold value of $\lambda^{(1)}_{\min}\approx -1.42$ and the existence of an upper branch. The profiles of two solutions obtained for the same value of $\lambda=1.5$  but on the different branches, upper and lower, are also shown in Figs.~\ref{fig:ccurve-point-b}~and~\ref{fig:ccurve-point-c} respectively.


\subsubsection{Multiple peak solutions}

The numerical continuation can also be applied in other areas of the solution space to trace different branches, whose presence is indicated by previous studies~\cite{Veveakis2014,Regenauer-Lieb2013a} who noted the existence of several peaks in the solution for higher values of $\lambda$. For the sake of demonstrating the existence and behavior of those solutions, we arbitrarily limit ourselves to solutions with up to seven peaks for ${\lambda<500}$.

For all given numbers of peaks ${(2 \leq  p \leq 7)}$, we apply the same procedure and identify manually a couple of solutions for values of $\lambda$ close to $500$ before running the continuation algorithm. Our procedure may not capture of all possible branches in the parameter space. Therefore, potentially even richer solutions exist. While the peak is always located at the center for single-peak solutions, as shown in Figure~\ref{fig:C-curve-1peak}, locations of the peaks do vary with $\lambda$ when there exists more than one peak. 
%

\begin{figure}[h!]
    \centering
    \includegraphics[width=0.9\textwidth]{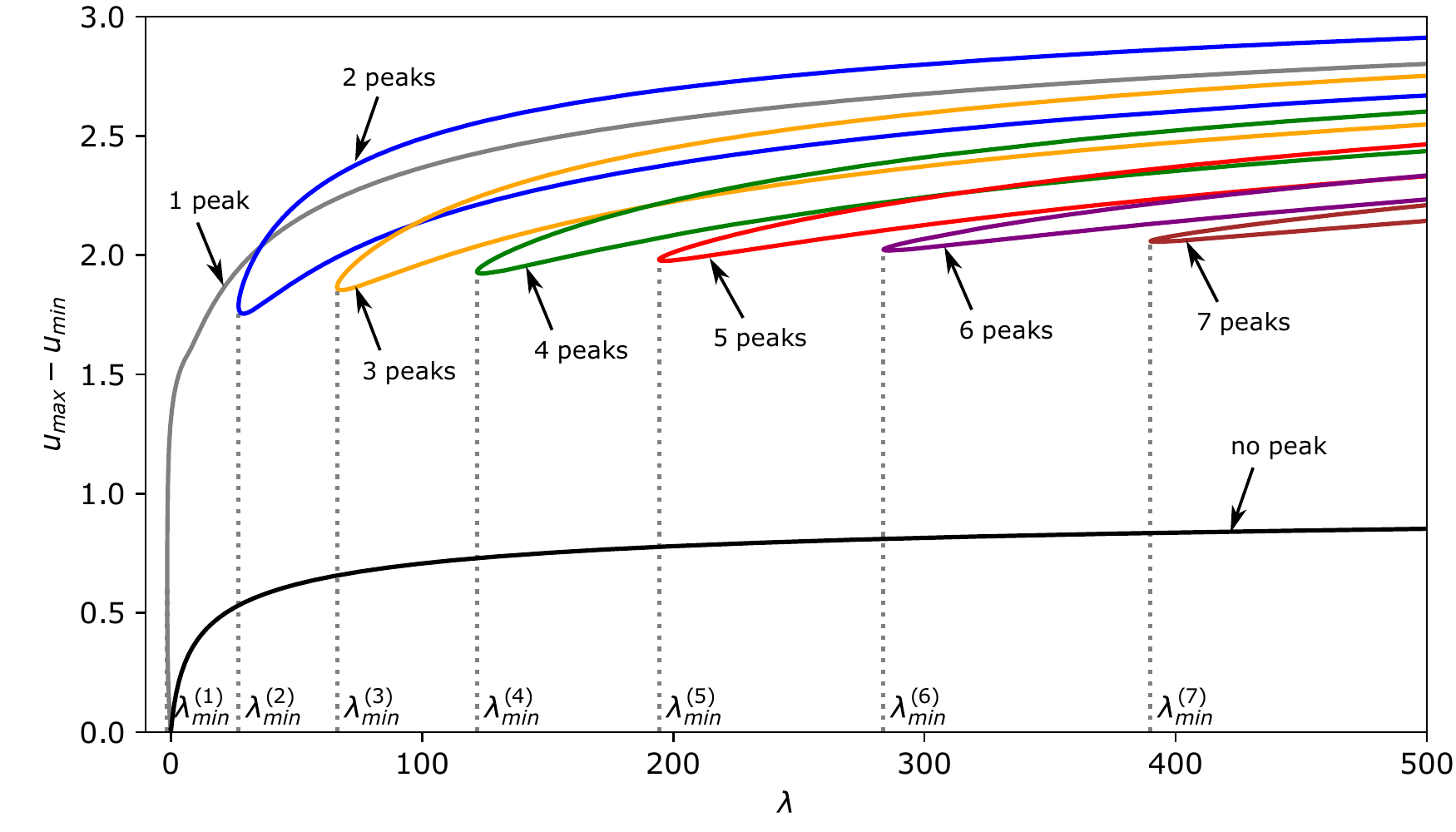}
    \caption{Map of the seven continuation curves obtained for $\lambda \leq 500$. All curves, for $p$ peaks, display a C-shape, marked by a minimum value $\lambda^{(p)}_{\min}$, as shown in Figure~\ref{fig:C-curve-1peak} for $p=1$.}
    \label{fig:space_map}
\end{figure}

Figure~\ref{fig:space_map} shows the corresponding results obtained for $\lambda \leq 500$, plotting the solution amplitude for each point of the curves. Similarly to the single-peak branch ($p=1$), all branches for $p\geq 2$ display a C-shape, indicating the existence of a lower threshold $\lambda^{(p)}_{\min}$ limiting the validity region of each branch. Each C-shape curve also indicates that there are at least two possible profiles for any given number $p$ of peaks for any admissible value of $\lambda \geq \lambda^{(p)}_{\min}$, which differ by the shape of the solution in general, with the same maximum value but different spacings between peaks.

\begin{figure}[ht!]
    \centering
    \begin{subfigure}[c]{0.75\textwidth}
        \centering
        \includegraphics[width=\textwidth]{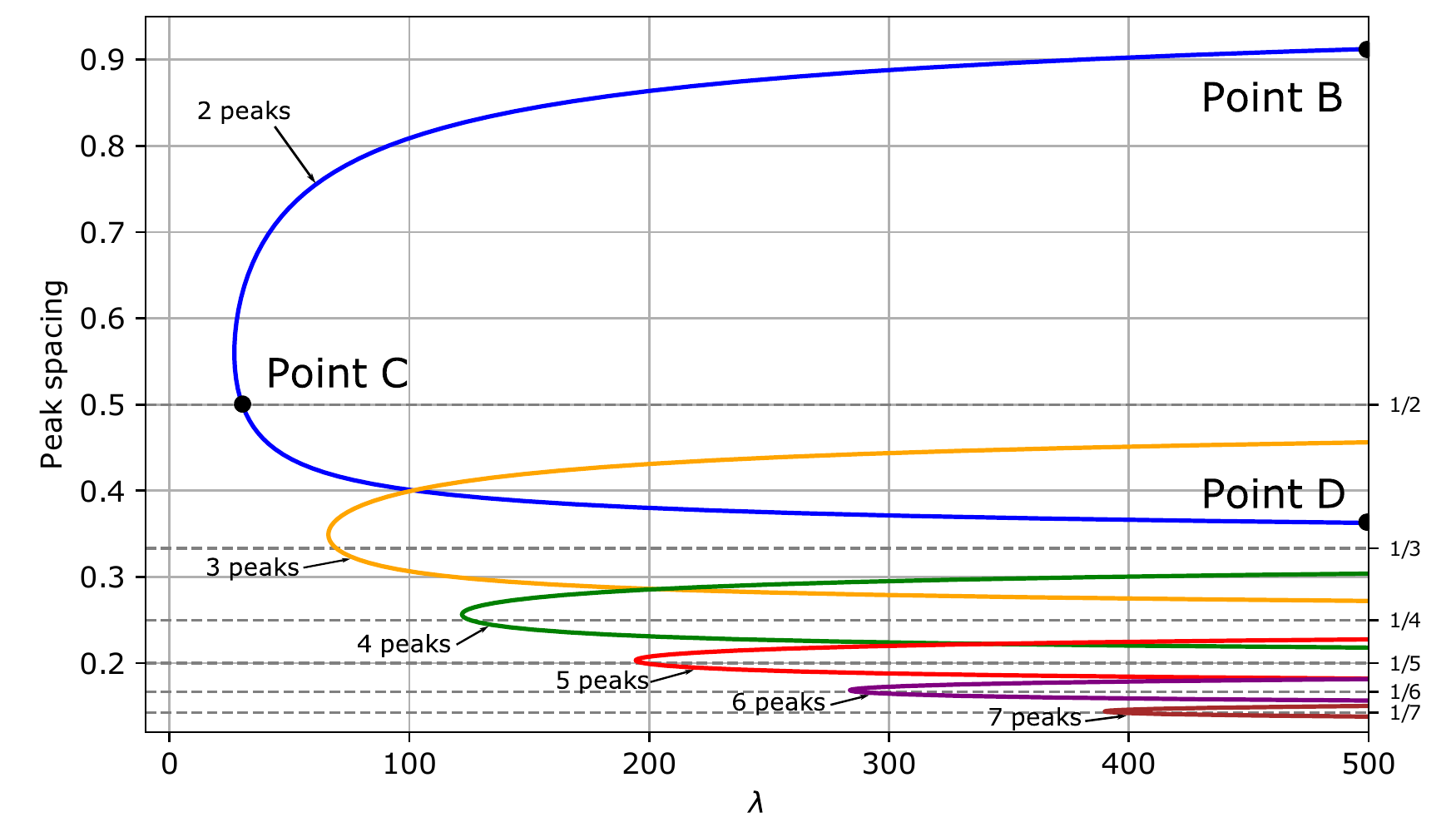}
        \caption{Spacing between peaks for branches with $p$ peaks (${2\leq p \leq7}$), with two values of $\lambda$ highlighted: ${\lambda^{(2)}_{\min}}$ (marking the lowest value of $\lambda$ on the C-curve) and ${\lambda^{(2)}_\text{eqd}}$ (value of $\lambda$ at point C, where ${spacing=1/p}$).}
        \label{fig:spacing-a}
    \end{subfigure}
    \begin{subfigure}[c]{0.26\textwidth}
        \centering
        \includegraphics[width=\textwidth]{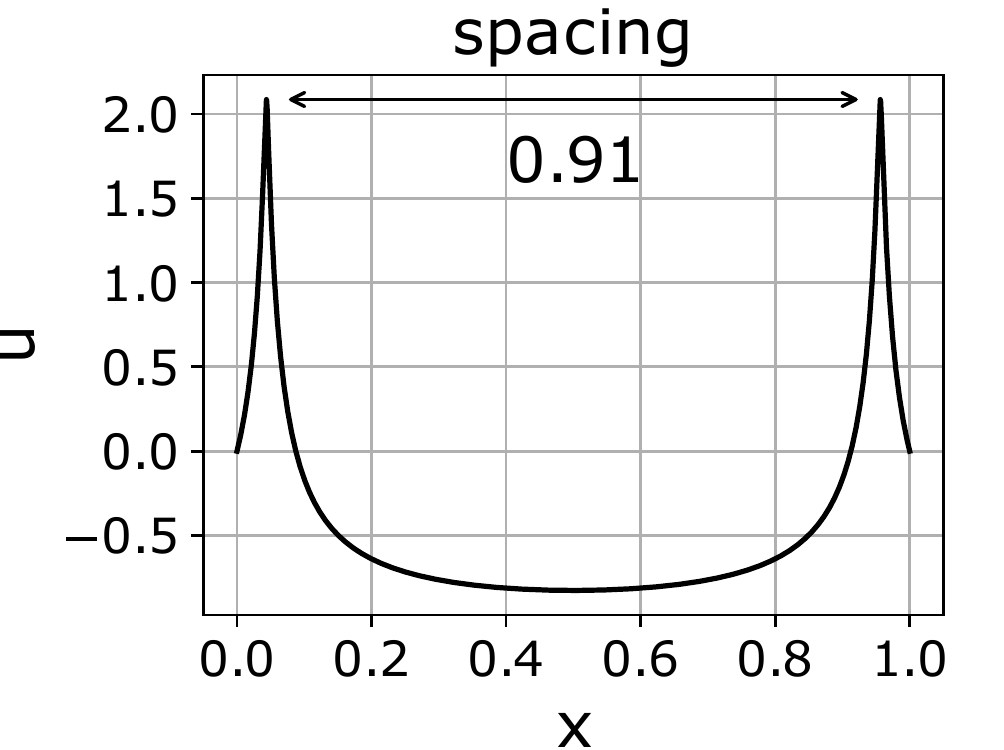}
        \caption{Point B: far-peaks sol.}
        \label{fig:spacing-b}
    \end{subfigure}
    \begin{subfigure}[c]{0.26\textwidth}
        \centering
        \includegraphics[width=\textwidth]{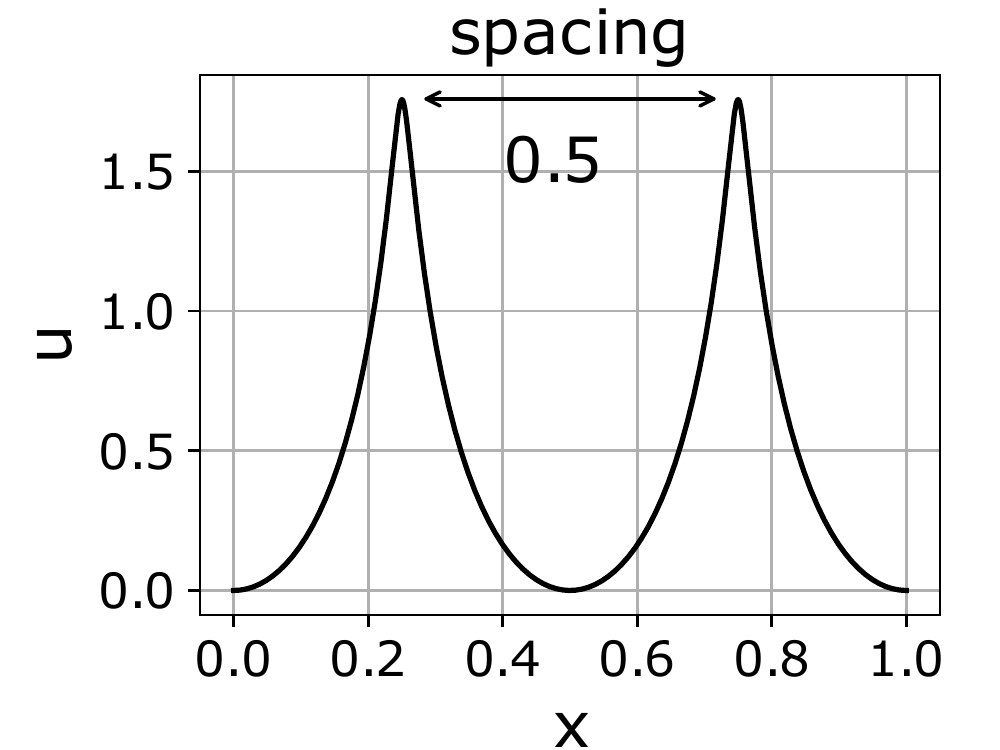}
        \caption{Point C: equidistant-peaks sol.}
        \label{fig:spacing-c}
    \end{subfigure}
    \begin{subfigure}[c]{0.26\textwidth}
        \centering
        \includegraphics[width=\textwidth]{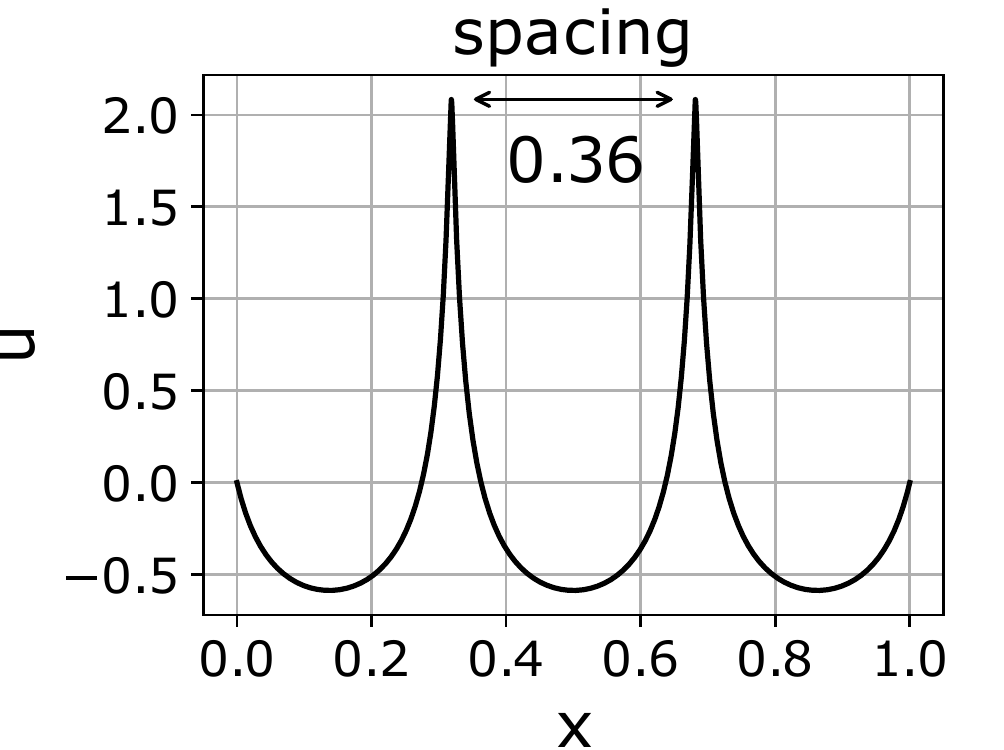}
        \caption{Point D: close-peaks sol.}
        \label{fig:spacing-d}
    \end{subfigure}
    \begin{subfigure}[c]{0.75\textwidth}
        \centering
        \includegraphics[width=\textwidth]{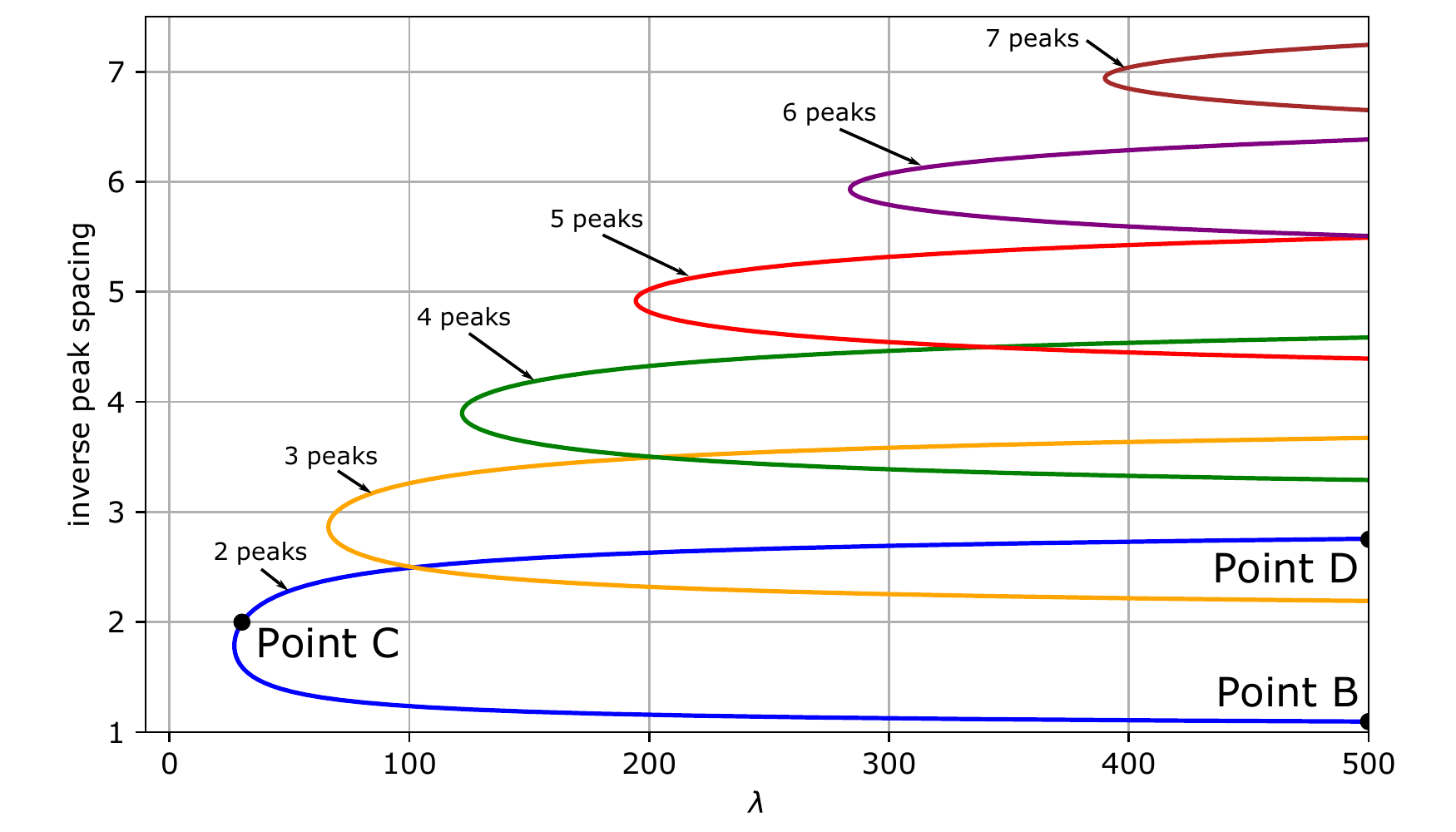}
        \caption{Inverse spacing plot, highlighting for each C-curve with $p$ peaks the turning point  (marked by point C for ${p=2}$) where ${1/spacing=p}$,  as well as the asymptotic behavior ${\lim_{\lambda\to\infty} spacing=p+1}$ on the upper branch (point B) and  ${\lim_{\lambda\to\infty} spacing=p+1}$ on the lower branch (point D). }
        \label{fig:spacing-e}
    \end{subfigure}
        \caption{Spacing maps as a function of $\lambda$ for every branch of the $p$-peak solution (${2 \leq p \leq  7}$).}
        \label{fig:peak_spacing}
    \end{figure}

We compute the distance between peaks by post-processing, noting that the regularity of all solutions provides constant spacing for the various peaks on any given profile. Figure~\ref{fig:peak_spacing} shows the spacing evolution (and its inverse) as a function of $\lambda$ for every branch of the $p$-peak solution for $2 \leq p \leq  7$, with all branches following the same pattern. For example, the two-peak solution on Figures~\ref{fig:spacing-b}-\ref{fig:spacing-d}. Figure~\ref{fig:spacing-a} shows the real spacing, Figure~\ref{fig:spacing-e} the inverse of that spacing, appears in Figures~\ref{fig:spacing-b}-\ref{fig:spacing-d} three specific profiles at points B, C and D highlighted on Figures~\ref{fig:spacing-a}~\&~\ref{fig:spacing-e}.

The equidistant spacing of $1/p$  (for ${x \in [0,1]}$) for all $p$ peaks is obtained for ${\lambda=\lambda^{(p)}_\text{eqd}}$, slightly larger than ${\lambda^{(p)}_{\min}}$, with the corresponding point on the C-curve for $p=2$ highlighted as point C on Figures~\ref{fig:spacing-a}~\&~\ref{fig:spacing-e}. Figure~\ref{fig:spacing-c} shows the corresponding solution profile. This point C naturally separates the higher and lower branches, leading to point B (resp. D) with increasing (resp. decreasing) spacing. Figures~\ref{fig:spacing-e} shows the asymptotic behavior ${\lim_{\lambda\to\infty} \text{spacing}=p+1}$ on the upper branch (point B) and  ${\lim_{\lambda\to\infty} \text{spacing}=p+1}$ on the lower branch (point D). The three solution profiles at points B, C and D are shown in Figs.~\ref{fig:spacing-b}-\ref{fig:spacing-d} and highlight this spacing evolution across the two-peak branches. All other branches (for ${p>2}$) in Figures~\ref{fig:peak_spacing} display the same behavior. 

\subsubsection{Evolution of number of peaks with respect to $\lambda$}
\begin{figure}[ht!]
    \centering
    \includegraphics[width=0.475\textwidth]{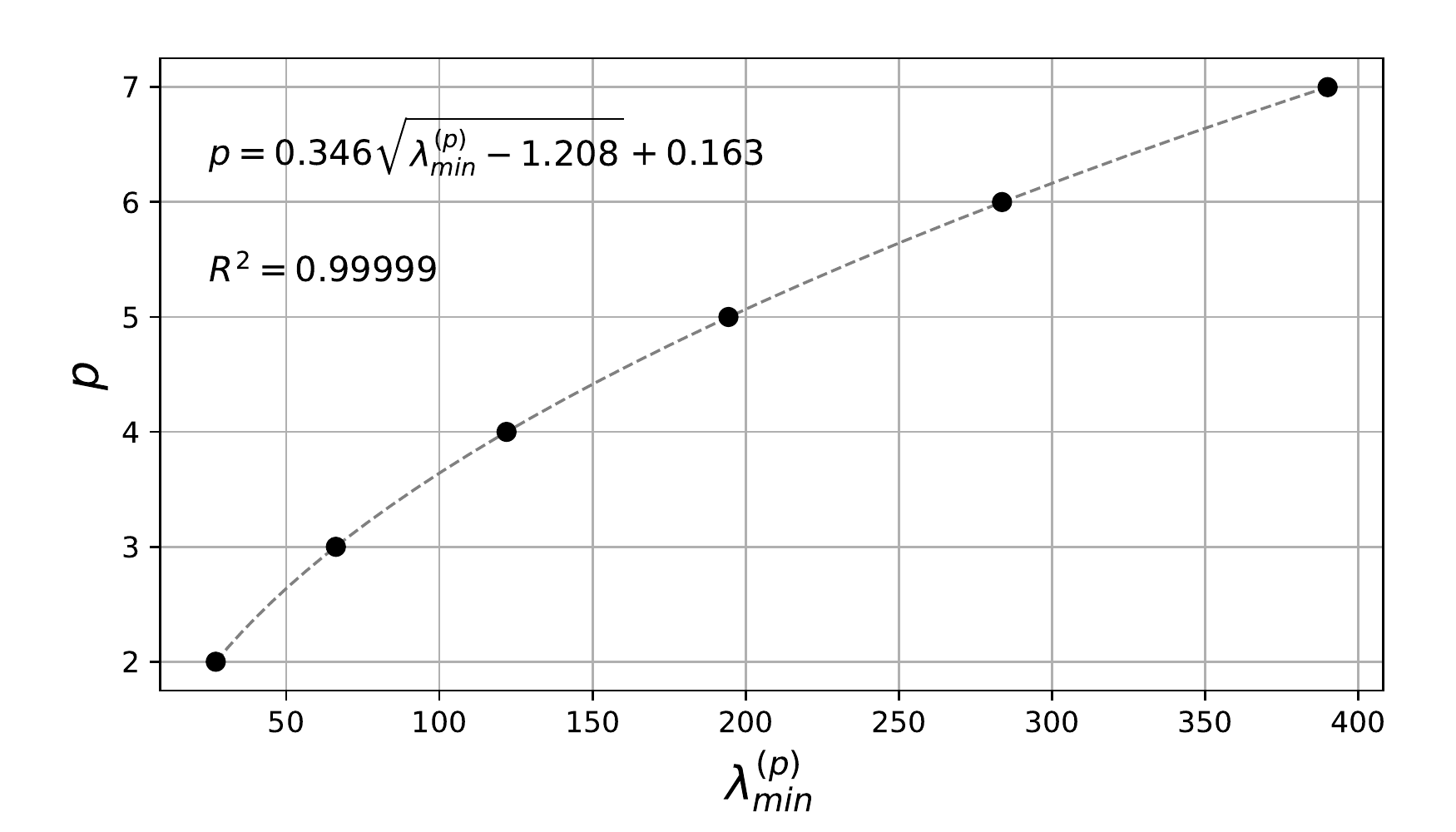}
    \includegraphics[width=0.475\textwidth]{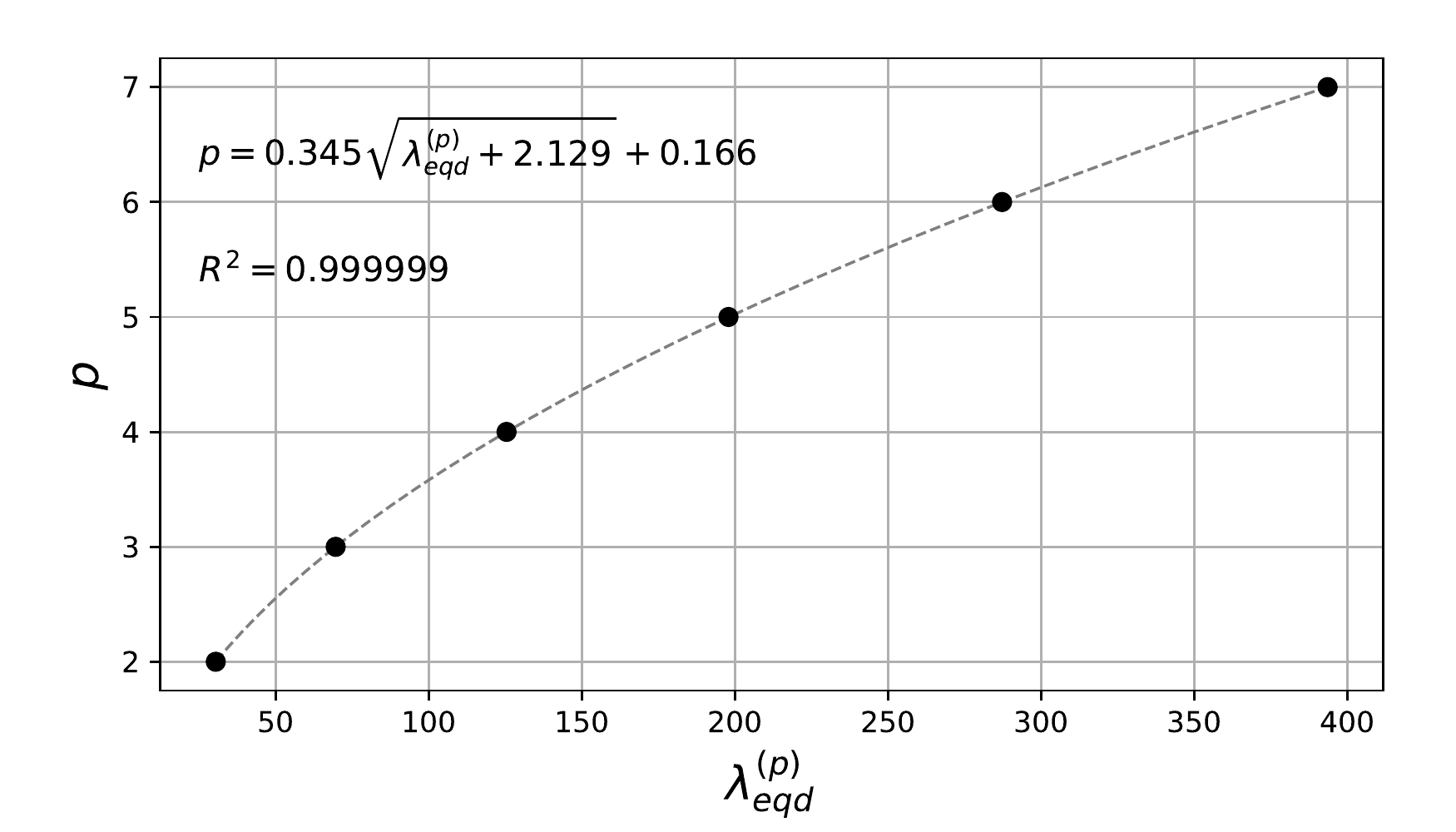}
    \caption{Evolution of $p$, the number of peaks, with respect to $\lambda^{(p)}_{\min}$ and  $\lambda^{(p)}_\text{eqd}$ as defined in Figure~\ref{fig:peak_spacing}.}
    \label{fig:sqrt_lambda}
\end{figure}

The identification of the natural turning point for ${\lambda=\lambda^{(p)}_\text{eqd}}$  on the lower branch of each C-curve, which is slightly different from the point marking the minimum possible of ${\lambda}$ ${(\lambda^{(p)}_\text{min})}$, indicates that both characteristic points are useful to refer to the ``start" of a C-curve. Table~\ref{tab:lambdas} lists the numerical values, with a precision of two decimals, of ${\lambda^{(p)}_\text{min}}$ and ${\lambda^{(p)}_\text{eqd}}$ for ${1\leq p \leq 7}$ as shown on Figure~\ref{fig:space_map}.
Figure~\ref{fig:sqrt_lambda} plots those evolutions of ${\lambda=\lambda^{(p)}_\text{min}}$ and ${\lambda=\lambda^{(p)}_\text{eqd}}$ with respect to $\lambda$. Fitting of those curves shows correspondence for both cases with the square-root relationship previously identified in~\cite{Regenauer-Lieb2013a} between $\lambda$ and the number of peaks $p$.

\begin{table}[ht!]
    \small
    \centering
    \caption{Values for each branch with $p$ peaks in Figure~\ref{fig:peak_spacing}, of ${\lambda^{(p)}_{\min}}$, the minimum value of $\lambda$ reached, and ${\lambda^{(p)}_\text{eqd}}$, the value of $\lambda$ for which the peaks are equidistant with ${spacing=1/p}$ (naturally, no such value exists for ${p=1}$).}
    \begin{tabular}{|c|c|c|c|c|c|c|c|} 
        \hline
        $p$ &1 & 2 & 3 & 4 & 5 & 6 & 7 \\
        \hline
        $\lambda^{(p)}_\text{min}$ &-1.42&26.99& 66.21&121.93&194.38&283.70&390.00\\
        $\lambda^{(p)}_\text{eqd}$ &--&30.30&69.48&125.32&197.84&287.23&393.59\\ 
        \hline
    \end{tabular}
    \label{tab:lambdas}
\end{table}

\section{Discussion and conclusions}
\label{sec:discussion}

Regarding the chemical effects in the physical formulation, following the work from~\cite{Alevizos2017} provides a bounded solution of the original problem of cnoidal waves in solids~\cite{Veveakis2015,Regenauer-Lieb2013a}. We solve the resulting nonlinear equation using an adaptive stabilized finite element framework~\cite{calo2019} for a wide spectrum of scenarios. We adopt a numerical continuation analysis to map the influence of the diffusivity ratio $\lambda$ in (\ref{eq:cnoidal_chem}) on the solution behavior, providing new insights on this recently described volumetric instability. Figure~\ref{fig:C-curve-1peak} shows that we can retrieve both the classical Terzaghi's isochrone (Figure~\ref{fig:ccurve-point-b}) or a peak-type solution (Figure~\ref{fig:ccurve-point-c}) for positive values of $\lambda$ and not only inside a specific range of integer numbers, as implied in~\cite{Veveakis2015,Regenauer-Lieb2013a}. This outcome represents a starting point for further experimental investigations to check whether this response can be observed in laboratory tests.

We also recover the previously identified square root relationship of the number of peaks in the solution with $\lambda$~\cite{Veveakis2015,Regenauer-Lieb2013a} and analyze in detail the solution space, (cf., see Figure~\ref{fig:space_map}). The C-shape nature of all $p$-peak solutions identifies minimum thresholds ($\lambda^{(p)}_{\min}$) above which those solutions exist and Figure~\ref{fig:sqrt_lambda} shows that $\lambda^{(p)}_{\min}\propto\sqrt{\lambda}$. Figure~\ref{fig:space_map}, however, also shows that all $p$-peak solutions exist in a stable manner past the $\lambda$ value of $\lambda^{(p+1)}_{\min}$, where the $(p+1)$-peak solution appears. This result has important consequences for the physical understanding of the problem as the number of bands measured on natural observations cannot be automatically correlated with $\lambda$ as~\cite{Veveakis2015} states. Furthermore, Figure~\ref{fig:peak_spacing} shows that the spacing between bands follows itself a more complex pattern than initially thought, since a $p$-peak solution exists for all values of $\lambda\geq\lambda^{(p)}_{\min}$ on two branches (see Figure~\ref{fig:space_map}), with the spacing between peaks increasing with $\lambda$ (asymptotically towards $1/(p-1)$) on the lower branch and decreasing with $\lambda$ (asymptotically towards $1/(p+1)$) on the upper branch. Figure~\ref{fig:peak_spacing} also shows that the natural turning point for each C-curve with $p$ peaks corresponds to the solution profile with all peaks equidistant at a spacing of $1/p$, obtained for $\lambda=\lambda^{(p)}_\text{eqd}$. The number of peaks $p$ follows as well a square root relationship concerning those ($\lambda^{(p)}_\text{eqd}$) values.

All simulations assume Dirichlet boundary conditions of (\ref{eq:cnoidal_chem}). Thus, the most appropriate boundary conditions that can match the natural observations are still open~\cite{Veveakis2015,Veveakis2014}. We deploy more advanced bifurcation analysis tools to complete the bifurcation map and identify all branches in the parameter space, including unstable ones, along with their characteristic points. Nonetheless, the surprisingly rich information obtained through introducing an appropriate numerical scheme to a simple generalization of the theory of consolidation, suggests that future works using this scheme in more elaborate elasto-viscoplastic formulations could enhance the mechanical solution with additional modes of localization stemming from the volumetric part of the plastic increment.

\section*{Acknowledgments}
This research was partially supported by the Australian Government through the Australian Research Council's Discovery Projects funding scheme (projects  DP170104550, DP170104557). This publication was also made possible in part by the CSIRO Professorial Chair in Computational Geoscience at Curtin University and the Deep Earth Imaging Enterprise Future Science Platforms of CSIRO. Additional support was provided by the European Union's Horizon 2020 Research and Innovation Program of the Marie Skłodowska-Curie grant agreement No. 777778, and the Mega-grant of the Russian Federation Government (N 14.Y26.31.0013). At Curtin University, The Institute for Geoscience Research (TIGeR) and by the Curtin Institute for Computation, kindly provide continuing support. MV acknowledges support by the DE-NE0008746-DoE project.

\appendixtitleon
\appendixtitletocon
\begin{appendices}
    
\renewcommand{\thesection}{A}
\numberwithin{equation}{section}

\section{Physical model: Compaction bands in saturated media}\label{sec:model}\label{AppxA}

For completeness, this section briefly recapitulates the formulation of the physical model behind~\eqref{eq:cnoidal_chem}, presented more in detail in~\cite{Alevizos2017}. We consider a one-dimensional representative elementary volume (REV) of porous material under compression in the $z$ direction. In this approach, the material is taken as homogeneous and all material properties are therefore constant. The sample of height $H$, under constant loading $p'_n$ at its boundaries, is considered already past its limit of elasticity and we track its mean effective stress $p'$ using the framework of overstress viscoplasticity by Perzyna~\cite{Perzyna1966}. Using Terzaghi's definition of effective stress $p = p' + p_f$, with $p$ the mean stress, taken positive in compression, and $p_f$ the pore pressure, we can express the momentum balance in the $z$ direction as
\begin{equation}
    \label{eq:mom_bal}
    \frac{\partial p'}{\partial z}=-\frac{\partial p_f}{\partial z}.
\end{equation}

Internal mass transfer is allowed between the solid and fluid phases through chemical reactions of dissolution/precipitation, which can be homogenized as a single effective reaction (see~\cite{Alevizos2017,Law2006}) written generically as
\begin{equation}
    \label{eq:chem_reaction}
    AB_{(solid)} \rightleftharpoons A_{(solid)} + B_{(fluid)}.
\end{equation}
Defining the solid and fluid phase densities as 
\begin{subequations}
    \label{eq:phase_densities}
    \begin{align}
        & \rho_{1}=(1-\phi) \rho_{s}, \\
        &\rho_{2}=\phi \rho_{f},
    \end{align}
\end{subequations}
where $\phi$ denotes the porosity, and $\rho_s$ (resp. $\rho_f$) the solid (resp. fluid) density, the mass balance equations of the solid and fluid phases can be written as 
\begin{subequations}
    \label{eq:mass_balance_phases}
    \begin{align}
        & \frac{\partial \rho_{1}}{\partial t}+\frac{\partial\left(\rho_{1} v_{z}^{(1)}\right)}{\partial z}=j, \\
        &\frac{\partial \rho_{2}}{\partial t}+\frac{\partial\left(\rho_{2} v_{z}^{(2)}\right)}{\partial z}=j,
    \end{align}
\end{subequations}
with $j$ the mass rate of fluid produced by the chemical reaction (\ref{eq:chem_reaction}) and $v_{z}^{(1)}$  and $v_{z}^{(2)}$ the velocities of phases 1 and 2 respectively. Combining (\ref{eq:mass_balance_phases}) with Darcy's law for the filter velocity $\phi\left(v_{k}^{(2)}-v_{k}^{(1)}\right)=-\frac{k}{\mu} \frac{\partial p_{f}}{\partial z}$ (with constant permeability $k$ and fluid viscosity $\mu$) leads to the mass balance equation for the solid-fluid mixture~\cite{Veveakis2014}
\begin{equation}
    \label{eq:mass_balance}
    -\frac{k}{\mu} \frac{\partial^{2} p_{f}}{\partial z^{2}}+\dot{\epsilon}_{V}=j\left(\frac{1}{\rho_{f}}-\frac{1}{\rho_{s}}\right),
\end{equation}
where $\dot{\epsilon}_{V}$ denotes the volumetric strain rate. Combining (\ref{eq:mom_bal}) and (\ref{eq:mass_balance}), we obtain
\begin{equation}
    \label{eq:5}
    \frac{k}{\mu} \frac{\partial^{2} p'}{\partial z^{2}}+\dot{\epsilon}_{V}=j\left(\frac{1}{\rho_{f}}-\frac{1}{\rho_{s}}\right).
\end{equation}
The volumetric strain rate is then decomposed into its elastic and (visco)plastic components, $\epsilon^{e}_V$  and $\epsilon^{vp}_V$, with the latter expressed through a typical power law rheology~\cite{Kohlstedt1995}, under isothermal and overstress assumptions
\begin{equation}
    \label{eq:strain_rate}
    \dot{\epsilon}_{V}=\dot{\epsilon}^{e}_V+\dot{\epsilon}^{vp}_V=-\frac{\dot{p}'}{K} - \dot{\epsilon}_{n}\left[\frac{p^{\prime}-p_Y^{\prime}}{p_{n}^{\prime}-p_Y^{\prime}}\right]^{m},
\end{equation}
where $K$ is the bulk modulus, $m$ the stress exponent, $p_Y$ the yield value, $p'_n$ the loading boundary conditions for $z\in\{0,H\}$ and $\dot{\epsilon}_{n}$ the corresponding loading strain rate. The negative signs match the sign convention of positive stresses in compression.
Using the overstress definition $\bar{p}=p'-p'_Y$, with $p'_Y$ constant, along with (\ref{eq:strain_rate}), (\ref{eq:5}) becomes
\begin{equation}
    \label{eq:7}
    \frac{k}{\mu} \frac{\partial^{2} \bar{p}}{\partial z^{2}}-\frac{1}{K}\frac{\partial\bar{p}}{\partial t} - \dot{\epsilon}_{n}\left[\frac{\bar{p}}{\bar{p}_{n}}\right]^{m}=j\left(\frac{1}{\rho_{f}}-\frac{1}{\rho_{s}}\right).
\end{equation}

All variables can be normalized
\begin{subequations}
    \label{eq:normalisation}
    \begin{align}
        & \sigma = \frac{\bar{p}}{\bar{p}_{n}} ,\\
        & \tau = \frac{k K}{\mu H^2} t,\\
        & z^*=  \frac{z}{H},
    \end{align}
\end{subequations}
and following~\cite{Alevizos2017}, the rate of fluid production follows an Arrhenius relationship with a dependence on mean pressure of the activation enthalpy. 
Assuming pressure-enhanced precipitation, it can be expressed as
\begin{equation}
    \label{eq:j}
    j = -A \mathrm{e}^{\beta \sigma}.
\end{equation}
(\ref{eq:7}) then gets rewritten in dimensionless form as
\begin{equation}
    \label{eq:cnoidal_1}
    \frac{\partial\sigma}{\partial\tau} = \frac{\partial^{2} \sigma}{\partial z^{*2}}  - \lambda \sigma^{m}  + \eta \mathrm{e}^{\beta \sigma},
\end{equation}
with  $\lambda=\frac{\mu H^2 \dot{\epsilon}_{n}}{k \bar{p}_{n}}$ and  $\eta=\frac{A \mu H^2}{k \bar{p}_n}\left(\frac{1}{\rho_{f}}-\frac{1}{\rho_{s}}\right)$. 
Dropping the asterix and considering the stationary case $\partial / \partial t = 0$, we recover~\eqref{eq:cnoidal_chem}. 
    
\end{appendices}

\bibliography{cnoidal}
\end{document}